# Elucidating microstructural influences on fatigue behavior for additively manufactured Hastelloy X using Bayesian-calibrated crystal plasticity model

Ajay Kushwaha [a], Eralp Demir [b], Amrita Basak [a]*

[a] Department of Mechanical Engineering, The Pennsylvania State University, University Park, PA 16802
[b] Department of Engineering Science, University of Oxford, OX1 3PJ, UK
*communicating author

**Abstract**

Crystal plasticity (CP) modeling is a vital tool for predicting the mechanical behavior of materials, but its calibration involves numerous (>8) constitutive parameters, often requiring time-consuming trial-and-error methods. This paper first proposes a robust calibration approach using Bayesian optimization (BO) to identify optimal CP model parameters under fatigue loading conditions. Utilizing cyclic data from additively manufactured Hastelloy X specimens at 500 °F, the BO framework, integrated with a Gaussian process surrogate model, significantly reduces the number of required simulations. A novel objective function is developed to match experimental stress-strain data across different strain amplitudes. Results demonstrate that effective CP model calibration is achieved within 75 iterations, with as few as 50 initial simulations. Sensitivity analysis reveals the influence of CP parameters at various loading points on the stress-strain curve. The results show that the stress-strain response is predominantly controlled by parameters related to yield, with increased influence from backstress parameters during compressive loading. In addition, the effect of introducing twins into the synthetic microstructure on fatigue behavior is studied, and a relationship between microstructural features and the fatigue indicator parameter is established. Results show that larger diameter grains, which exhibit a higher Schmid factor and an average misorientation of approximately 42°±1.67°, are identified as probable sites for failure. The proposed optimization framework can be applied to any material system or CP model, streamlining the calibration process and improving the predictive accuracy of such models. Insights into key microstructural features influencing fatigue behavior enables precise control over material properties, facilitating the design of more durable materials under fatigue loading.

## 1. Introduction

Crystal plasticity (CP) models are widely used for predicting the deformation behavior of single crystal and polycrystalline materials by considering the underlying grain morphology and crystallographic texture [1], [2]. These models play a key role in simulating material responses under various loading conditions, offering insights into microstructural phenomena such as strain localization and texture evolution [2], [3]. CP models are especially useful in fatigue analysis, where they help predict component life, likely fatigue initiation sites, and the interaction of microstructural features like twins and pores [4], [5], [6]. However, the accuracy and predictive capability of CP models depend heavily on the proper calibration of their constitutive laws, which typically involve a large number of parameters ($> 8$). As the complexity of the CP model increases, so does the number of parameters, making calibration to experimental data both computationally intensive and challenging. Failure to properly calibrate the model can result in inaccurate predictions of material behavior and fatigue life predictions.

The available constitutive models for material behavior can generally be classified into two main categories: phenomenological models and physics-based models [1]. Within these categories, various slip models such as the hyperbolic sinh law [7], power law [1], and Orowan equation [3] are commonly used. Similarly, several strain hardening models, including the linear evolution of dislocations [8], Voce [9], and Kocks-Mecking [10] models are employed. For kinematic hardening, models like the Armstrong-Frederick [11] and Ohno-Wang [12] models are widely applied. Each of these models requires a specific set of parameters, many of which cannot be directly obtained through experimental techniques. To identify and calibrate parameters, various experimental techniques, such as nanoindentation, micropillar compression [13], strain mapping using digital image correlation [14], and high-energy X-ray diffraction microscopy [15] have been used. However, the calibration process using these methods is resource-intensive and requires significant experimental effort. Given the complexity and cost of these experimental approaches, macroscopic stress-strain data from tensile or cyclic tests is often preferred for parameter identification [16], [17].

The inverse optimization problem for parameter identification in CP modeling can be addressed using various methods, including trial-and-error approaches [18], gradient-based optimization techniques like the Levenberg-Marquardt algorithm [19], and gradient-free methods such as the Nelder-Mead algorithm [20]. However, these techniques can be inefficient and are often susceptible to converging on local minima, particularly due to the high dimensionality and non-linearity inherent in CP models [21]. An increasingly popular alternative for calibrating CP parameters is the use of Genetic algorithms (GAs) [6], [22], [23], [24], which avoid many of the limitations associated with gradient-based and gradient-free methods. GAs are more robust in exploring the parameter space, making them well-suited for complex optimization problems in CP modeling [22].

Skippon et al. [24] attempted to determine CP parameters for an elastoplastic self-consistent (EPSC) polycrystalline model of Zircaloy-2 using GA, finding the results comparable to manual calibration. However, GA struggled to identify satisfactory hardening parameters related to tensile twinning, likely due to the chosen fitness function. They argued that combining a better CP model with GA would enhance the performance of the algorithm. Later, Prithivirajan and Sangid [6] used GA to identify CP parameters for cyclic loading of IN718, achieving a good fit. They then used it to further estimate the critical pore size for additively manufactured (AM) IN718. Bandyopadhyay et al. [22] explored the uncertainty in mechanical response due to variations in CP parameters after optimizing parameters using GA. The results suggested that variability in stress, plastic strain and plastic strain energy density values may exist due to uncertainty in CP parameters. Sedighiani et al. [23] applied response surface methodology (RSM) as a surrogate for CP simulations alongside GA for optimization. They successfully identified parameters for both dislocation-density-based and phenomenological CP models and quantified the role of the underlying single crystal parameters on the deformation behavior. However, their work relied on synthetic monotonic stress-strain data rather than experimental data. Building on this, Savage et al. [25] introduced a multi-objective GA to optimize the EPSC framework, incorporating both mechanical and microstructural evolution data for improved accuracy. They found that a good model fit can be achieved solely based on stress-strain response. However, this approach of using multi-objective GA was computationally expensive due to the large number (238,680) of simulations required to converge to an optimal solution for a total of 26 parameters.

While GAs are effective for parameter identification in CP simulations, they often suffer from significant computational expense, requiring numerous evaluations of the objective function, which can be resource-intensive due to the complexity of the simulations involved. In contrast, Bayesian optimization (BO) offers a more efficient approach by utilizing a probabilistic model to estimate the objective function and guide the search for optimal parameters. This method strategically selects points in the parameter space that are expected to yield substantial improvements, thereby reducing the number of simulations needed. As a result, BO enhances both the efficiency and accuracy of calibrating CP parameters, effectively balancing exploration and exploitation. It has been successfully applied to other applications like optimizing process parameters for laser-directed energy deposition (L-DED) [26] and design parameters for pin-fin arrays to achieve superior efficiency [27].

Tran and Lim [28] used the BO algorithm to calibrate phenomenological CP models for three different material systems (stainless steel 304L, Tantalum, and Cantor high-entropy alloy), demonstrating its ability to identify optimized parameters for monotonic loading conditions. Sun and Wang [29] enhanced the BO method with a Monte Carlo Markov Chain (MCMC) based strategy to efficiently calibrate CP model parameters, significantly reducing the search domain and number of simulations required. They demonstrated this approach's effectiveness on the magnesium alloy ZEK100 and concluded that the total number of simulations could be reduced to a hundred. Kuhn et al. [30] applied a Gaussian Process (GP) surrogate with BO to optimize CP parameters for two backstress models in high-strength steel for fatigue applications. They compared BO with an evolutionary algorithm and other derivative-free methods such as Multilevel Coordinate Search and Stable Noisy Optimization by Branch and FIT, finding BO consistently outperformed the alternatives. However, their study optimized only three parameters, taking hardening and slip parameters from previous studies, and focused on a single fatigue cycle, which did not capture hardening behavior across multiple cycles.

After optimizing the CP model, it can be used to understand the effect of microstructural features on low cycle fatigue (LCF) performance. Studies have shown that microstructural features play a vital role in strain localization both for traditional and AM materials. Grain boundaries

(GBs), specifically twin boundaries (TBs), and microstructural anomalies such as pores and inclusions play a role in the nucleation of microcracks in nickel-based superalloys for AM materials. Slip accumulation during cyclic loading leads to strain localization around these features which further leads to crack initiation [31]. Liu et al. [32] concluded that the GBs were the preferred locations for crack initiation and propagation due to the oxidation near the GB surface at high temperature high cycle fatigue tests. They concluded that twins with 60° misorientation affect the high cycle fatigue crack propagation. Abuzaid et al. [33] conducted fatigue experiments for Hastelloy X with digital image correlation to understand the role of GBs. They observed high strain around boundaries especially TBs, however they showed that not all TBs behave the same, some allow the slip transmission while others block the transmission. The boundaries blocking the slip transmission led to crack initiation. To conclude, microstructure attributes like GBs and annealing twins play a vital role in mechanical properties especially fatigue performance of AM parts.

There have been very few CP studies conducted on understanding the fatigue behavior of laser-powder bed fusion (L-PBF) AM Hastelloy X. Aburakhia et al. [34] developed a CP model for L-PBF Hastelloy X to understand how mechanical anisotropy and the evolution of internal strains are affected by changing the L-PBF parameters for compression tests. Pilgar et al. [35] developed a microstructure-sensitive model using CP-based on a Fast Fourier Transform (FFT) solver to predict the life of Hastelloy X specimens fabricated with different orientations. Cyclic and accumulated plastic work were used as fatigue indicator parameters (FIP) to predict life for both strain and stress-controlled tests, respectively. However, the annealing twins were not included in simulations, which as discussed earlier affects the fatigue crack initiation and therefore needs to be incorporated in the simulations. Zhao et al. [36] showed that including annealing twins in synthetic microstructures can further enhance fatigue life predictive capabilities. However, the authors didn't fully explore what factors are leading to the enhanced prediction. Although efforts have been made to use CP modeling to study specimens produced via L-PBF, microstructural factors contributing to fatigue failure of Hastelloy X specimens manufactured via L-PBF have been sparsely studied.

A thorough review of the literature highlights several key challenges. These include the high computational cost associated with calibrating CP models, the need for a deeper understanding of the effects of introducing TBs in synthetic microstructures, and the identification of critical microstructural features that influence the LCF performance of Hastelloy X produced via L-PBF process. To address these issues, in this study, an automated GP-based BO framework is proposed for the inverse identification of CP parameters, featuring a novel objective function. Applied to a strain gradient-based CP model for L-PBF Hastelloy X, this framework aims to accurately replicate the macroscopic stress-strain behavior over two consecutive cycles based on experimental data. The study not only quantifies the optimal number of initial simulations required for effective optimization but also incorporates SHapley Additive exPlanations (SHAP)-based sensitivity analysis to investigate the influence of various parameters on the cyclic stress-strain response. This detailed analysis provides insights into the contributions of individual parameters, enhancing the understanding of CP models and their influence on material behavior. By quantifying the minimum number of simulations needed to achieve reliable results, the research offers practical guidelines for reducing computational costs while maintaining high accuracy. Furthermore, this study discusses the role of TBs during fatigue loading conditions and identifies critical microstructural attributes.

The paper is structured into four sections. Following this introductory section, section 2 outlines the material characterization, generation of the 3D microstructure, CP model equations, and the BO framework. Section 3 then presents the key findings from the parameter optimization and sensitivity analysis. Twins are observed in L-PBF Hastelloy X and hence in section 4, impact of introducing twins in synthetic microstructures is explored. Finally, the paper concludes with section 5, summarizing the study's main findings and their implications.

## 2. Methods
### 2.1 Experimental
#### 2.1.1 Specimen manufacturing and fatigue testing

The experimental results are based on the work of Pal et al. [37]. The details of the experiments conducted are briefly summarized here. The specimens were fabricated using a Concept Laser M2 system at Solar Turbines Incorporated in San Diego, USA. Gas-atomized Hastelloy X powder with a nominal composition of Cr 20.5-23%, Fe 17-20%, Mo 8-10%, and Ni (balance) was used. The nominal powder size distribution was 15 to 45 µm. To fabricate the specimens, a volumetric energy density of 70 J/mm$^3$ was employed with a layer thickness of 35 µm. After the printing process, all specimens underwent a hot isostatic pressing (HIP) cycle, applying typical industrial settings. Following the HIP treatment, two nickel braze thermal exposures were applied to each specimen, which were subsequently machined for characterization and testing.

Strain-controlled fatigue tests were conducted on an MTS® 370.10 thermomechanical fatigue test system, with specimen designs conforming to the ASTM E606 standard [38]. Data from two different strain amplitudes: 0.5% and 0.75%, applied at a loading (R) ratio of -1 (i.e., fully reversed loading conditions) was used from the LCF tests for machined specimens. The strain values were applied to the specimens in the form of a triangular waveform, with the strain rate ranging from 0.01 and 0.02 s$^{-1}$. All the tests were conducted under an isothermal condition of 500 °F (260 °C), which corresponds to the operating temperature of fuel injectors in land-based gas turbine engines where Hastelloy X components are typically utilized.

#### 2.1.2 Microstructure characterization

For this study, three specimens are sectioned longitudinally to the print direction from the grip section for metallographic analysis. The specimens are prepared according to ASTM E3-11 [39] (sectioning, mounting, coarse grinding, fine grinding, and then polishing using diamond abrasive). The specimens are sectioned, mounted in Bakelite, and subjected to wet grinding with

SiC paper, gradually reducing the grit size from #120 to #800. Then the specimens are polished using 3-µm and 1-µm diamond solutions and 0.05-µm colloidal alumina suspension until a mirror-like finish is achieved. A Thermo Fisher Scientific Apreo S scanning electron microscope equipped with Oxford Instruments Electron Backscatter Diffraction (EBSD) detector is utilized for microstructure characterization. For grain size characteristics and crystallographic orientations, EBSD measurements are performed on polished specimens using a scanning step size of 2 µm. Oxford Instruments AztecCrystal software is used to construct the maps of GBs and band contrast (Figure 1(a)) with relative Σ3 {111} coincidental site lattice (CSL) boundaries as shown in Figure 1(b). The Σ3 boundaries are defined with a misorientation deviation of ±8.66° from 60° following Brandon's criterion [40]. Figure 1(c) illustrates the distribution of the grain equivalent circle diameter in the microstructure. The histogram reveals a significant number of smaller grains, with only a few larger grains present in the distribution. This skewed distribution suggests a predominantly fine-grained microstructure.

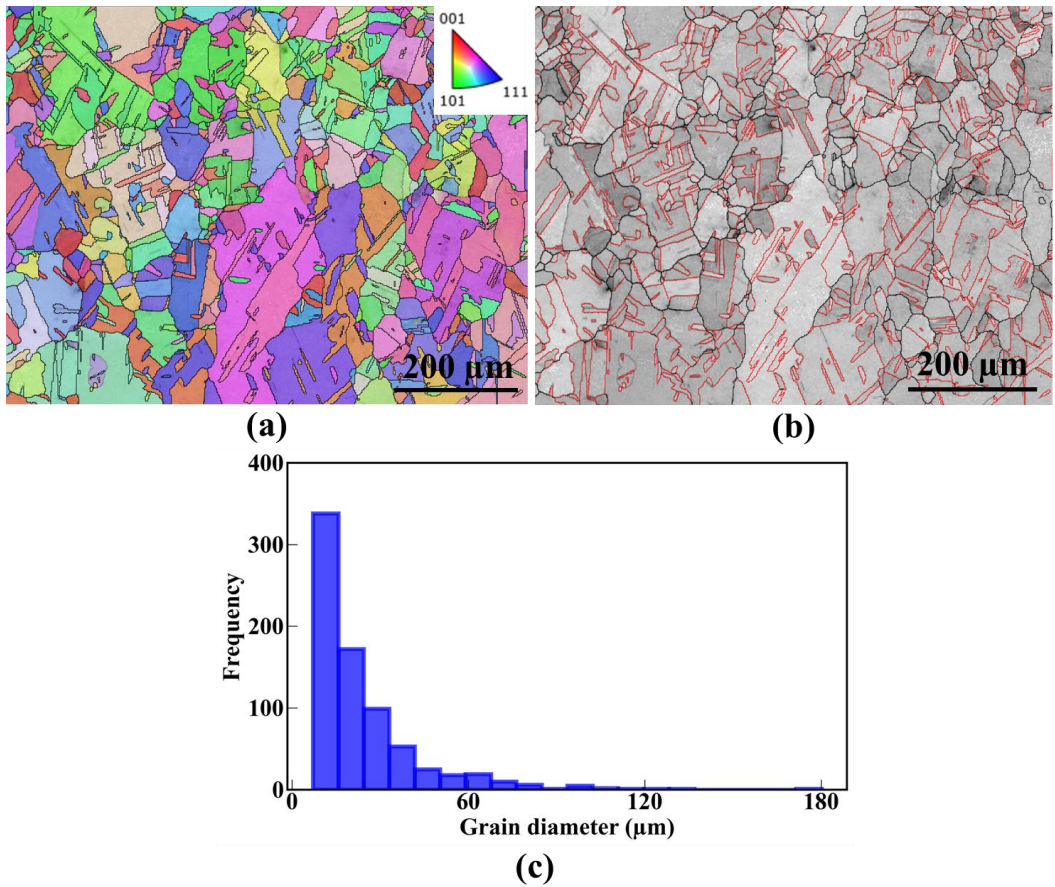

**Figure 1.** (a) EBSD micrograph of a representative specimen fabricated via L-PBF. (b) A map of low angle (gray color lines) and high angle grain boundaries (black color lines) along with Σ3 {111} CSL boundaries (red color lines). (c) Histogram plot of equivalent circle diameter of the grains in the map.

EBSD characterization is performed on three specimens and the mean equivalent diameter is determined to be 23.37 µm. The twin fraction (Σ3 boundaries) is found to vary between 0.62-0.73. Table 1 provides a summary of the grain size characteristics and Σ3 CSL boundaries for three different specimens, showing that the microstructure is largely consistent across all the specimens. The mean grain diameters are similar, ranging from 23.16 µm to 23.55 µm, indicating uniformity in grain size. The minimum grain diameter remained constant at 7.14 µm, while the maximum diameter shows some variation, with specimen 2 having a smaller maximum (117.46 µm) compared to specimens 1 and 3 (around 180 µm), suggesting occasional larger grains in the latter two specimens. The standard deviation values, between 17.98 and 19.46 µm, are relatively high, compared to the mean grain diameter (23.37 µm), indicating that the grain size distribution is broad, with a mix of smaller and significantly larger grains in the microstructure. This suggests that while the average grain size is consistent across the specimens, there are substantial differences in individual grain sizes within each specimen, leading to a heterogeneous grain structure. The ASTM grain size, around 7.4-7.5, further supports the overall uniformity across the specimens. An ASTM grain size of 7.4 indicates that the grains are relatively fine, meaning that the microstructure has a significant number of smaller grains. The proportion of Σ3 CSL boundaries, while slightly variable, remains high across all specimens.

**Table 1.** Grain size characteristics and percentage of Σ3 CSL boundaries for three sectioned specimens obtained through EBSD

| Specimen# | Mean diameter (µm) | Minimum diameter (µm) | Maximum diameter (µm) | Standard deviation (µm) | ASTM 2627 Grain size | Σ3 CSL boundaries (%) |
|---|---|---|---|---|---|---|
| 1 | 23.55 | 7.14 | 180.16 | 19.2 | 7.4 | 69.2 |
| 2 | 23.39 | 7.14 | 117.46 | 17.98 | 7.5 | 61.9 |
| 3 | 23.16 | 7.14 | 178.43 | 19.46 | 7.4 | 72.5 |

**2.2 Computational**
**2.2.1 Synthetic microstructure generation**

Following characterization, synthetic 3D microstructures are generated using the open-source software Dream3D (version 6.5.171) [41]. Statistical microstructural data such as grain size, orientations, and twin-length fractions are extracted from the EBSD data. Grain size is represented by the diameter of an equivalent circle, and orientations are in Bunge Euler angles notation. Since EBSD data provides only 2D grain sizes (equivalent circle diameter), 3D grain sizes (equivalent sphere diameters) are obtained by multiplying 2D grain size by a factor of $4/\pi$ [22], [42]. DREAM.3D's synthetic building filters are utilized to generate the microstructure, which involves inputs such as 3D grain sizes, orientation data, desired microstructure volume, and resolution (voxel size). To include twins, the 'Insert Transformation Phases' filter is used. Twins are inserted into the microstructure in a randomly chosen grain by the filter. Due to limitations with the software, representative volume elements (RVEs) are generated with a twin fraction ranging from 0.4 to 0.5, whereas EBSD maps show a twin fraction ranging from 0.62 to 0.73. Abaqus input files are generated by Dream.3D using linear 8-node brick hexahedral elements (C3D8). For this study, 10 RVEs are generated, each shaped as a cube with dimensions of 200 μm on each side. Each RVE contains around 300 grains, which is adequate for accurately representing the texture and capturing the homogenized stress-strain response at the macroscopic level [43].

### 2.2.2 Crystal Plasticity modeling

The strain gradient crystal plasticity model is incorporated in ABAQUS® (v. 2022) within a user-defined material subroutine OXFORD-UMAT available at [44]. A detailed summary can be found in [45]. The total deformation gradient ($F$) associated with the elastoplastic deformation of a solid body is multiplicatively decomposed as:

$$F = F^e F^p, \tag{1}$$

where, $F^e$ and $F^p$ are deformation gradients associated with the elastic and plastic parts, respectively [1]. Further, the plastic velocity gradient, $L^p$ is related to $F^p$ as:

$$L^p = \dot{F}^p F^{p-1} \tag{2}$$

The plastic velocity gradient, $L^p$ is written as a sum of the contributions from all slip systems. Since Hastelloy X is an FCC material, the sum is performed over all ⟨110⟩{111} slip systems as follows:

$$L^p = \sum_{a=1}^{N} \dot{\gamma}^a s^a \otimes n^a, \tag{3}$$

where, the vectors $s^a$ and $n^a$ are the slip direction and the slip plane normal, respectively, and $N$ is the number of slip systems which equals 12; $\dot{\gamma}^a$ is the shear rate for slip system $a$. A Hutchinson-type flow rule [46] is used as follows:

$$\dot{\gamma}^a = \dot{\gamma}_0 \left| \frac{\tau^a - \chi^a}{\tau_c^a} \right|^n sgn(\tau^a - \chi^a), \tag{4}$$

where, $\dot{\gamma}^a$ is the shear rate for slip system $a$ subjected to the resolved shear stress $\tau^a$ at a slip resistance $\tau_c^a$; $\dot{\gamma}_0$ is a reference shear rate, $n$ is the inverse strain rate sensitivity exponent and $\chi^a$ is backstress. Taylor's relation [47] is used to compute effective critical resolved shear stress $(\tau_c^a)_{eff}$.

$$(\tau_c^a)_{eff} = \tau_c^0 + CGb^a \sqrt{\rho_{for}^a} + \tau_c^a, \tag{5}$$

where, $\tau_c^0$, $C, G, b^a$, $\tau_c^a$ and $\rho_{for}^a$ represent the lattice friction, geometric factor, shear modulus, Burger's vector, strength due to statistical hardening, and forest dislocation density, respectively. Forest dislocations are obtained by the projection of total dislocations:

$$\rho_{for}^a = \xi_b^a \rho_{tot}^b, \tag{6}$$

where, $\xi_b^a$ is the forest projection of a dislocation with line direction, $l^b$ onto a slip system with slip plane normal $n^a$.

$$\xi_b^a = |n^a \cdot l^b| \tag{7}$$

The total dislocation density is expressed as:

$$\rho_{tot}^a = |\rho_{GND,s}^a| + |\rho_{GND,e}^a| + \rho_{SSD}^a, \tag{8}$$

where, $\rho_{GND,s}^a$ and $\rho_{GND,e}^a$, are screw and edge type of geometrically necessary dislocations (GND) densities and $\rho_{SSD}^a$ is statistically-stored dislocations (SSDs) density. SSDs are assumed to be constant as the applied strain considered in this study is small (<0.0075) and SSDs contribute more to the strength at the larger applied strains [48]. GND densities are calculated using the curl of plastic deformation gradient ($\boldsymbol{F^p}$) followed by singular value decomposition inversion as described in [47]. The influence of any slip system $b$ on the statistical hardening behavior of slip system $a$ is given by:

$$\dot{\tau}_c^a = h_{ab}|\dot{\gamma}^b|, \tag{9}$$

where, $h_{ab}$ is referred to as the hardening matrix,

$$h_{ab} = q_{ab}\left[h_0\left(1 - \frac{\tau_c^b}{\tau_s}\right)^m\right], \tag{10}$$

where, $h_0$, $m$, and $\tau_s$ are slip hardening parameters [2]. The parameter $q_{ab}$ is a measure of latent hardening; its value is taken as 1.0 for coplanar slip systems $a$ and $b$, and 1.2 otherwise [4]. The backstress $\chi^a$ that accounts for the kinematic hardening follows the nonlinear evolution rule with an Armstrong–Fredrick-type equation [11] and is given by:

$$\dot{\chi}^a = h\dot{\gamma}^a - h_d\chi^a|\dot{\gamma}^a|, \tag{11}$$

where, $h$ and $h_d$ are the direct hardening modulus and the dynamic recovery modulus, respectively. Effective plastic strain rate $\dot{p}$ [49] is calculated using the following equation:

$$\dot{p} = \left(\frac{2}{3}\boldsymbol{L^p}:\boldsymbol{L^p}\right)^{\frac{1}{2}} \tag{12}$$

To predict a potential site for crack initiation, a metric known as fatigue indicator parameter is calculated at each material point within the model that incorporates physical attributes and mechanisms related to fatigue damage. One such metric is plastic strain energy density ($W$), which corresponds to the dissipative energy per unit volume due to plastic deformation [50].

$$W = \sum_{a=1}^{N} \oint \tau^a \dot{\gamma}^a \, dt \tag{13}$$

The nine CP parameters $n$, $\tau_c^0$, $C$, $\rho_{SSD}^a$, $h_0$, $\tau_s$, $m$, $h$, and $h_d$ are obtained by calibrating the model with macroscopic stress-strain response using BO. Elastic constants are initially taken from [51] but then adjusted to match the experimental results. Finally, $C_{11} = 250\ GPa$, $C_{22} = 139\ GPa$ and $C_{44} = 70.2\ GPa$ are used. The CP simulations are run for three fully reversed loading-unloading cycles for optimization. The last two cycles from the simulations are compared with two stable consecutive cycles from the experiments to capture the macroscopic response. During simulations, normal displacements on three mutually orthogonal adjacent surfaces of the RVE are restricted, and normal displacement is specified on another face, as shown in Figure 2, to mimic the uniaxial loading condition. XSYMM denotes boundary condition (BC) which is $u_x = R_y = R_z = 0$. Similarly, YSYMM denotes $u_y = R_x = R_z = 0$ and ZSYMM denotes $u_z = R_y = $

$R_x = 0$, here $u_i$ is displacement given in the $i$ direction ($i$ = x, y, or z) and $R_j$ is rotation around $j$ axis ($j$ = x, y, or z), respectively.

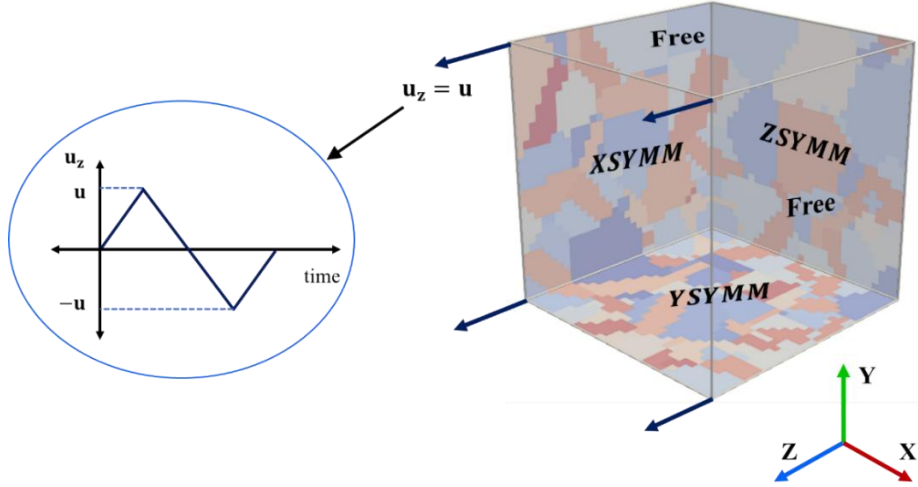

**Figure 2.** Boundary conditions used for CP simulations. Displacement (u) in triangular waveform is applied along z direction.

### 2.3 Bayesian Optimization

Bayesian Optimization (BO) is an advanced technique used to optimize black-box functions that are expensive to evaluate [52]. In this study, BO is employed to identify the set of CP parameters by iteratively selecting the most promising parameter sets for evaluation that would capture the stress-strain response of the material under uniaxial fatigue loading. The key idea behind BO is to construct a surrogate model, which guides the search for the next best solution, allowing for efficient exploration and exploitation of the parameter space.

#### 2.3.1 Gaussian process surrogate

The surrogate model used in this study is Gaussian process (GP) regressors. GPs are well-suited for modeling complex, non-linear functions and provide both a mean prediction and an estimate of uncertainty (variance) for any given input [53]. This uncertainty estimate is crucial for guiding the optimization process. A GP assumes that a collection of random variables has a joint Gaussian distribution. Given a set of input parameters $X = \{x_1, x_2, \ldots, x_n\}$ and corresponding outputs $y = \{y_1, y_2, \ldots, y_n\}$, the outputs are assumed to have a multivariate Gaussian distribution, typically expressed as:

$$y \sim \mathcal{N}(m(x), k(x, x')), \quad (12)$$

where, $\mathcal{N}$ implies a Gaussian distribution, and $y$ is drawn from $\mathcal{N}$ with a mean function m(x), which predicts the average response for the input $x$, and a covariance function $k(x, x')$. Covariance functions are crucial in GPs, as they determine the shape of the prior and posterior distribution of the model. In this study, the commonly used Matérn 5/2 kernel is employed as the covariance function, due to its flexibility and smoothness when modeling complex functions [26]. The goal is to predict the function value at new points in the parameter space using this probabilistic model. Since there are not many CP studies available for Hastelloy X, a wide range of parameter space (Table 2) is chosen based on studies of similar type of Ni-alloy, IN718 [4], [6], [22], [34], [35] and some preliminary understanding of the parameters. The GPyOpt library [54] is used to implement BO, leveraging the GP surrogate model to efficiently search for the optimal parameters. All computations are performed on a node with 128 GB of memory and 48 cores.

**Table 2.** Search domain to find optimum CP parameter for BO algorithm.

| Parameter | Description | Range |
| --- | --- | --- |
| $n$ | Strain rate sensitivity | [5, 25] |
| $\tau_c^0$ | Initial CRSS (MPa) | [20, 150] |
| $C$ | Geometric factor | [0.1, 0.5] |
| $\rho_{SSD}^a$ | SSD density ($\mu m^{-2}$) | [1, 100] |
| $h_0$ | Hardening rate (MPa) | [10, 1000] |
| $\tau_s$ | Saturation slip strength (MPa) | [50, 1000] |
| $m$ | Hardening exponent | [1, 15] |
| $h$ | Kinematic hardening modulus (MPa) | [1000, 80000] |
| $h_d$ | Recovery modulus | [0, 3000] |

### 2.3.2 Acquisition function: Expected Improvement (EI)

The acquisition function is a critical component of BO, guiding the selection of the next set of parameters to evaluate. In this work, the Expected Improvement (EI) acquisition function is employed. The EI function balances exploration and exploitation by selecting points that are expected to improve upon the current best-known value while considering the uncertainty of the predictions.

The EI function is defined as:

$$EI(x) = \mathbb{E}[max(f(x) - f(x^+), 0)], \qquad (13)$$

where, $f(x^+)$ is the best observed value so far, and $\mathbb{E}$ denotes the expectation under the GP posterior distribution. Points with high EI values are prioritized for evaluation, ensuring that the optimization process converges efficiently toward the global optimum.

### 2.3.3 Objective function

The objective function used in this study is defined as the root mean squared error ($\Delta\sigma_1$) between the experimentally observed stress ($\sigma_{exp}$) and homogenized stress ($\sigma_{sim}$) along the loading direction from CP simulations following previous studies [19], [22] as in Equation 14. The schematic is presented in Figure 3.

$$\Delta\sigma_1 = \sqrt{\left(\left(\frac{1}{n}\right) \times \sum_{i=1}^{n}(\sigma_{exp} - \sigma_{sim})^2\right)}, \quad (14)$$

where, n is the total number of points chosen on the stress-strain curve, which is set to 44 in this study as shown in Figure 3. Notably, the hardening between the two consecutive experimental cycles is less, so the points may seem to overlap. However, as shown in the inset, the points do not overlap. The goal is to minimize the error as described by Equation 14.

However, based on results as presented in section 3.3, an additional term is introduced into the objective function (Equation 15). This new term captures the average difference between the endpoints of two consecutive simulation cycles, with the aim of controlling the hardening behavior between these cycles as shown in Figure 3. The new objective function is described as:

$$\Delta\sigma = \sqrt{\left(\left(\frac{1}{n}\right) \times \sum_{i=1}^{n}(\sigma_{exp} - \sigma_{sim})^2\right)} + \lambda \times \frac{(\Delta\sigma_{max32} + |\Delta\sigma_{min32}|)}{2}, \quad (15)$$

where, $\Delta\sigma_{max32}$ is the difference between the stress at the strain end level during loading (i.e. maximum applied strain) between two consecutive cycles (i.e., 2nd and 3rd cycle) and $|\Delta\sigma_{min32}|$ is the absolute difference between the stress at the strain end level during compressive loading (i.e. maximum applied strain in reversed direction) between two consecutive cycles (i.e., 2nd and 3rd cycle) from simulations. Here, $\lambda$ is a weighing factor to control the influence of the new term introduced in the objective function. Three values of $\lambda$, i.e, 0, 0.5, and 1 are used in this study.

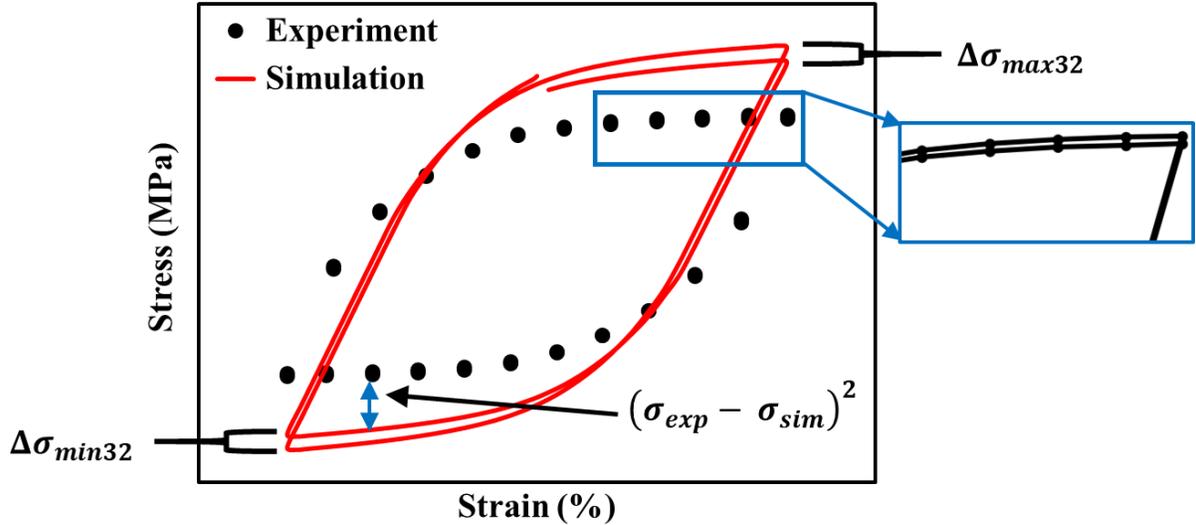

**Figure 3.** Schematic showing the calculation of objective function for example simulated curve and experimental data used.

### 2.3.4 Optimization process

The framework is depicted in Figure 4. The optimization process begins with an initial set of CP simulations, generated using a Latin Hypercube Sampling (LHS) design of experiments

(DOE) from the domain in Table 2. LHS is selected because it ensures a more uniform coverage of the multidimensional parameter space compared to simple random sampling [55]. Then the Abaqus input files are modified to run the simulations. The initial simulations are used to train the GP surrogate model. BO is then employed to iteratively suggest new parameter sets based on the EI acquisition function. For each new set of parameters, a CP simulation is conducted, and the results are used to update the GP model. This iterative process continues until the best possible result is found within the given computational budget, which is set to 75 iterations in this study. Since the surrogate model estimates the potential solution across the entire search space, BO can make predictions quickly without needing to evaluate the objective function at every point in the space (i.e., without having to run simulations for all possible combinations of CP parameters) unlike evolutionary algorithms.

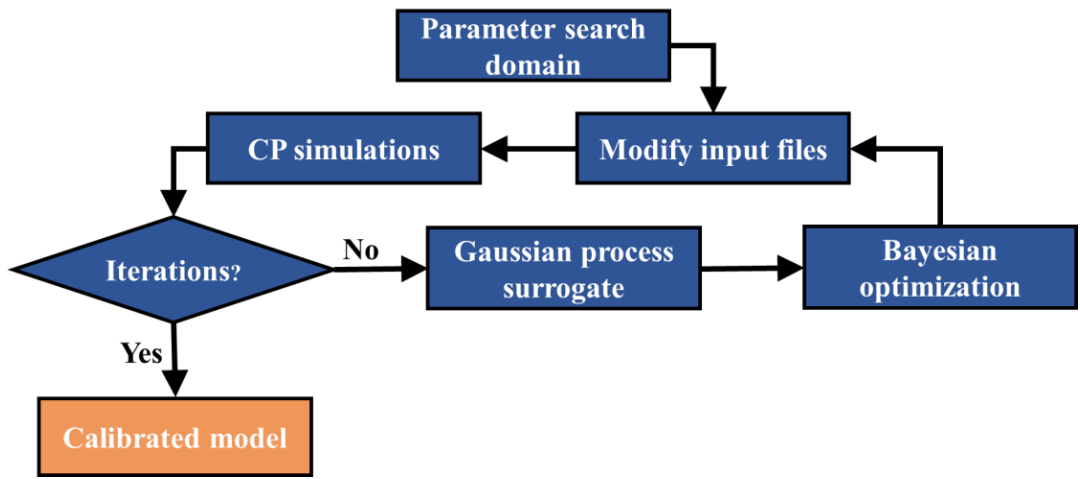

**Figure 4.** Flowchart depicting the framework for Bayesian optimization of CP parameters

**2.3.5 SHapley Additive exPlanations (SHAP) for sensitivity analysis**

SHapley Additive exPlanations (SHAP) is a game-theoretic approach to explain the output of machine learning models [56]. In the context of sensitivity analysis, SHAP values are used to assess the contribution of each input parameter to the predicted outcome, providing insights into the relative importance and influence of each variable. This provides a global understanding of the feature's importance across different regions of the input space, as well as local explanations for individual predictions.

In this study, after training a GP surrogate to model the relationship between CP parameters and homogenized stress along the cyclic stress-strain curve, SHAP values are computed to evaluate the sensitivity of each input parameter. By quantifying how much each parameter influences the stress predictions, SHAP provides an interpretable ranking of feature importance. This enables a more comprehensive understanding of which parameters drive the mechanical response at different stages of deformation. The sensitivity analysis is conducted using the initial dataset of 100 simulations, as well as the additional simulation data generated through BO for all three $\lambda$ values i.e., 0, 0.5, and 1. The combination of BO with GP surrogate model and sensitivity analysis using SHAP offers a comprehensive approach for efficiently identifying and understanding the key parameters in strain-gradient CP modeling.

## 3 Results
### 3.1 Grid convergence study

Mesh element size plays a crucial role in finite element simulations. While smaller sized elements or a higher number of elements can improve accuracy, they also increase the computational cost. To strike a balance between accuracy and efficiency, a mesh sensitivity study is typically conducted. This study helps determine the optimal mesh element size that provides accurate results without exhausting the computational resources. The mesh element size is adjusted in Dream3D using the resolution option in the 'Initialize Synthetic Volume' filter to obtain the desired meshed model. The mesh study is conducted for RVE of size 200 µm with approximately 300 grains. The element size is varied from 10 µm to 3 µm for tensile loading conditions using the same BCs as shown in Figure 2. The same CP parameters are used for all simulations. However, it's important to note that the mesh sensitivity study is conducted before the model calibration. Figure 5(a) shows the variation in results of the tensile stress-strain curve between various element sizes. It is observed that the percentage difference of stress values between 3 µm and 10 µm element size is about 1%. However, the time taken to run the simulation increases exponentially from 2 minutes for 10 µm element size to 45 minutes for 3 µm element size, as seen in Figure 5(b). Considering both the grain size distribution and computational time, the chosen element size is 6.25 µm.

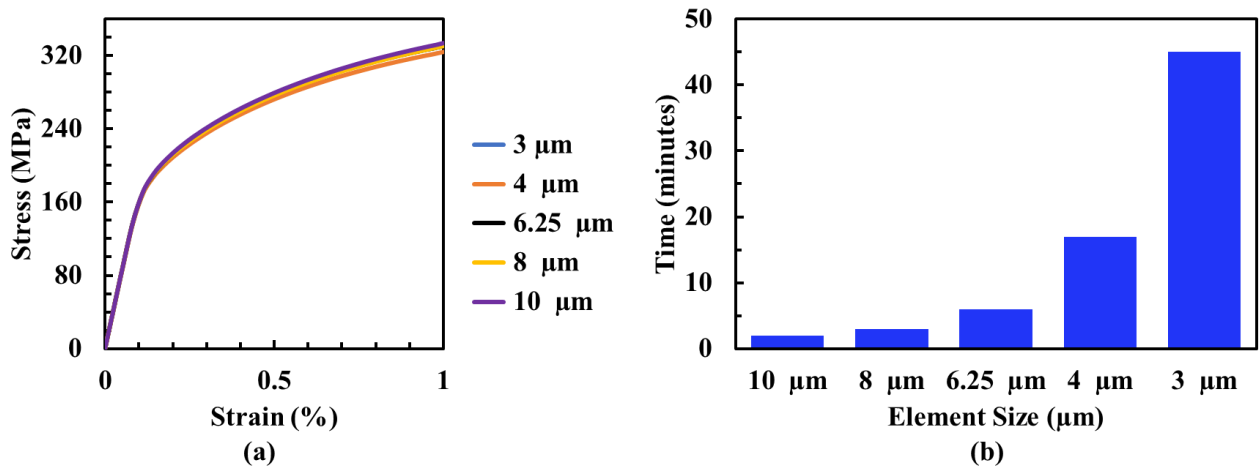

**Figure 5.** (a) Stress-strain response of RVE for different element sizes and (b) time taken to complete the simulation for each element size.

### 3.2 Performance of GP surrogate model

A well-performing GP surrogate model is essential for the success of BO, as it guides the search for optimal parameter sets by accurately predicting the objective function across the search domain. The performance of the surrogate model is evaluated using simulation data from 100 parameter sets generated through LHS. These sets are divided into a 70-30 split, with 70% used for training and 30% reserved for testing. The input to the GP model consists of the 9 CP parameters, and the output is corresponding Δσ, the error between the experimental and simulated stress-strain data as defined in Equation 15. The $R^2$ score is used to measure the model's accuracy in predicting Δσ. To ensure the robustness of the GP model, it is trained with three different values of the $\lambda$ factor (0, 0.5, and 1). This is done to confirm that introducing an additional term does not adversely affect the model's performance or the results of the subsequent optimization. The plot of predicted vs. true Δσ with the 95% confidence interval for predictions is shown in Figure 6. True values are shown as black dashed line, predicted values are represented by blue dots with red bars denoting twice the standard deviation of the prediction. The resulting $R^2$ scores are 0.93, 0.92, and 0.92 for $\lambda = 0$, 0.5, and 1, respectively, demonstrating that the GP model accurately predicts Δσ. All three plots appear similar, but there is a slight shift in the points to the right. This shift is subtle and difficult to distinguish because the second term's magnitude is small.

From Figure 6, it can be observed that the predicted points are symmetrically distributed around the black dashed line, with no evidence of systematic bias. This suggests that the model's predictions are reliable across the entire range of the data. The model demonstrates its ability to make accurate and unbiased predictions, as the error bars (red lines) indicate that uncertainty is relatively consistent and well-contained across all predictions. The close alignment of the predicted values with the black dashed line, along with high $R^2$ value, further supports the model's accuracy in predicting $\Delta\sigma$ for different $\lambda$ values. This confirms that the model performs well across various scenarios and is both robust and dependable.

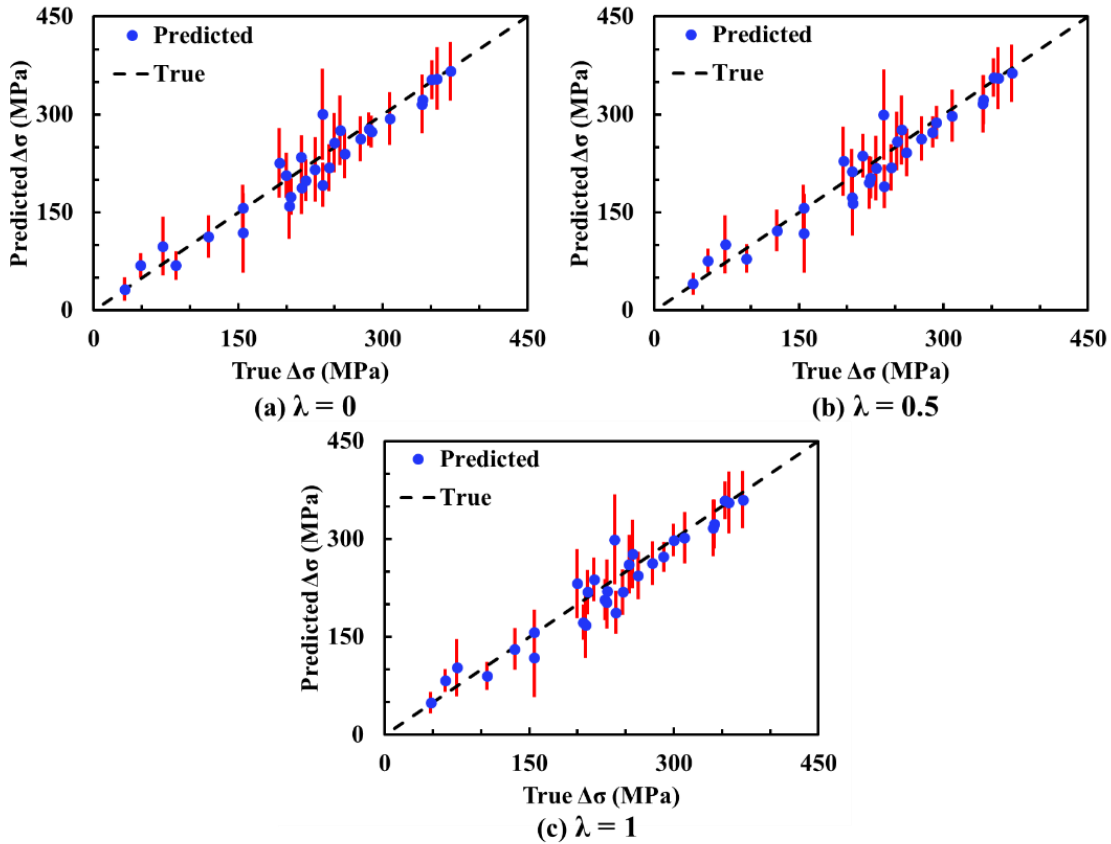

**Figure 6.** Predicted Δσ compared to true Δσ for (a) $\lambda = 0$, (b) $\lambda = 0.5$ and (c) $\lambda = 1$ with red bars representing twice the standard deviation of the prediction.

The initial number of parameter sets are chosen to be 100 at random to build the surrogate model. It is important to note that CP simulations are computationally expensive, and running such a large number of simulations can be a significant challenge. To address this, a study to assess the impact of varying the number of initial simulation data on the performance of BO is conducted. By testing the BO algorithm with different numbers of initial data points, this study identifies the minimum number of simulations required to achieve accurate CP model calibration.

### 3.3 Performance of BO and evaluation of objective function

Given the computational expense of CP simulations, the BO algorithm is initially tested for a strain amplitude of ±0.5%. The BO algorithm runs for 75 iterations, using a surrogate model built from an initial set of 100 parameter combinations. Figure 7 shows the convergence of the optimization, depicting the evolution of $\Delta\sigma$ (the objective function) over the BO iterations. Occasionally, $\Delta\sigma$ peaks at 500 MPa, which represents a penalty assigned to the algorithm when simulations fail for the parameters suggested by the BO algorithm. This penalty ensures that the algorithm avoids selecting parameters that may lead to failed simulations.

Overall, the BO algorithm achieves a low $\Delta\sigma$ value within a few iterations, indicating effective optimization. However, as seen in Figure 7(a), occasional peaks other than the penalty of 500 MPa occur due to the exploration phase of the algorithm. The algorithm identifies an optimal set of parameters with a final $\Delta\sigma$ of 17.95 MPa, an improvement over the initial 100 simulations, where the lowest $\Delta\sigma$ was 32.12 MPa. Despite this, Figure 7(b) shows that the simulated stress-strain curve does not align well with the experimental curve. The experimental curve, plotted for two consecutive cycles, shows nearly overlapping cycles, indicating minimal hardening. In contrast, the simulated stress-strain curve exhibits significant hardening between the two cycles. Therefore, to control this hardening behavior between cycles, an additional term is introduced into the objective function, weighted by a factor, $\lambda$. Figures 7(c), (d) and (e), (f) display the BO convergence plots and stress-strain curves for the optimal solutions obtained with $\lambda$ values of 0.5 and 1, respectively. The convergence behavior for $\lambda = 0.5$ and $\lambda = 1$ is similar to that of $\lambda = 0$, with a higher frequency of simulation failures observed for $\lambda = 1$. Least number of simulation failures are observed for $\lambda = 0.5$.

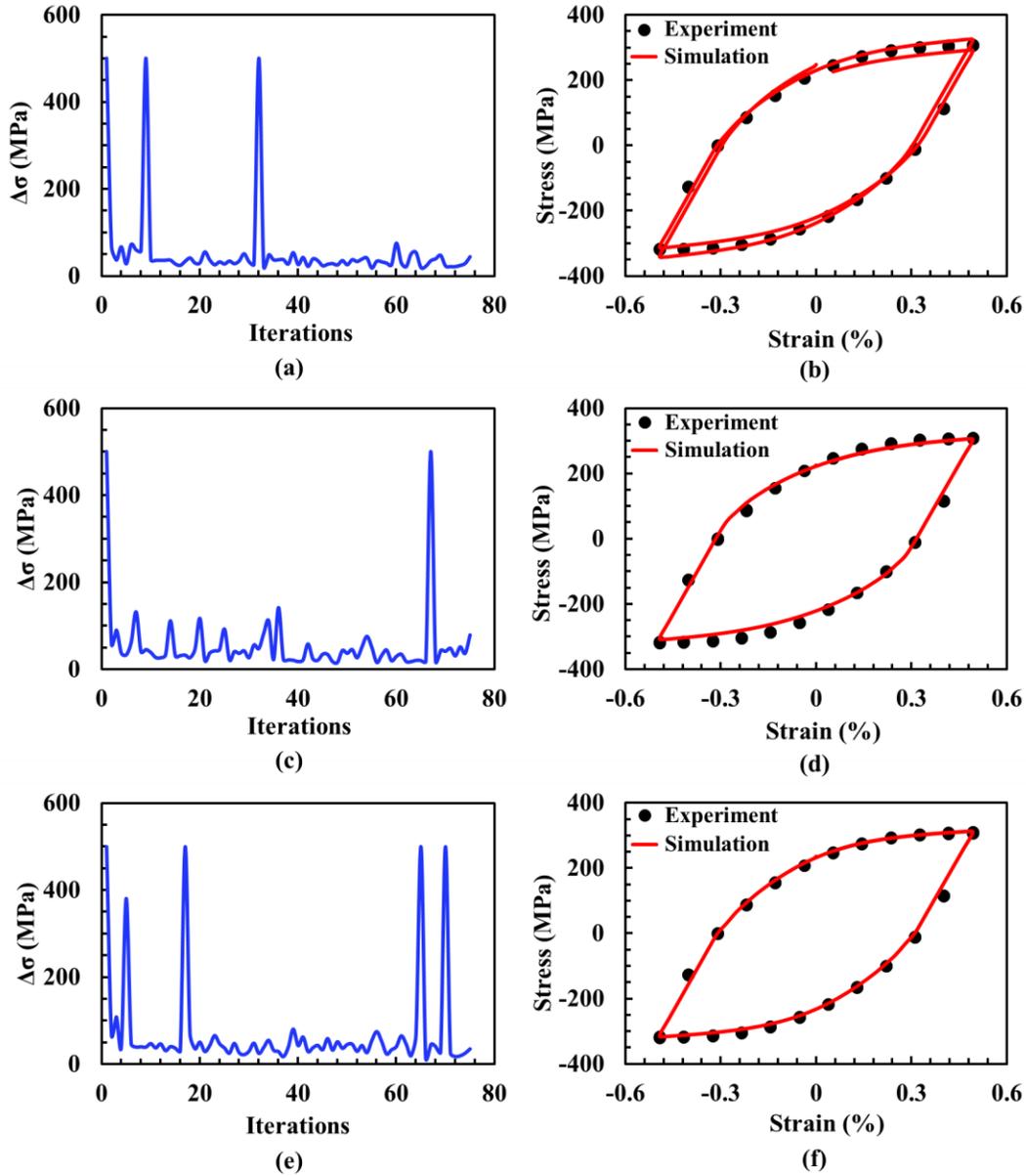

**Figure 7.** BO convergence plot and corresponding stress-strain response for optimized parameters for (a), (b) $\lambda = 0$, (c), (d) $\lambda = 0.5$ and (e), (f) $\lambda = 1$.

From Figures 7(d) and (f), it is evident that introducing $\lambda$ reduces the hardening response between the two simulated cycles. To better quantify the effect of $\lambda$ on hardening behavior, $\Delta\sigma_{max32}$ is plotted against the BO iterations. Figure 8 shows that the average value of $\Delta\sigma_{max32}$ decreases with increasing $\lambda$. However, if $\lambda$ is set too high, it may overemphasize the second term of the objective function, reducing the overall accuracy of the simulated stress-strain response.

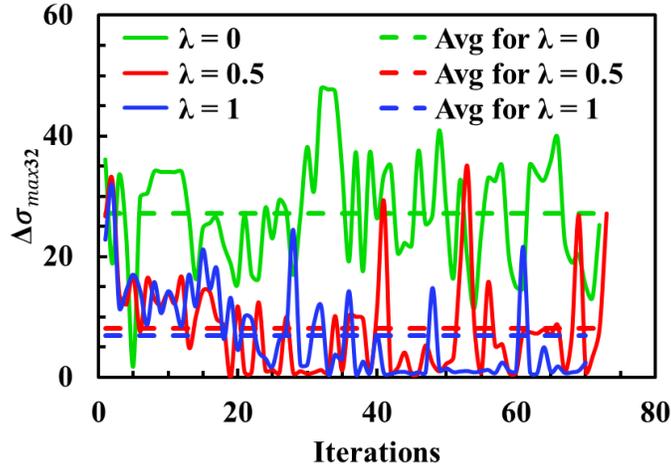

**Figure 8.** Evolution of the difference between the stress at the maximum strain during loading between two simulated consecutive cycles, $\Delta\sigma_{max32}$, over BO iterations for different $\lambda$ values

### 3.4 Assessing BO Performance with limited simulation data

After evaluating the BO algorithm using initial 100 prior simulations for the ±0.5% strain condition, its performance assessed with a reduced number of initial simulations. This analysis explores how the algorithm behaves when fewer simulations are available. The algorithm is run with initial 100, 50, 25, and 0 simulations. The initial simulations of 50 and 25 datapoints are taken from initial set of 100 by removing the best performing, i.e., simulations that have low $\Delta\sigma$ values. Figure 9 summarizes the results for each case. Figure 9(a) shows the number of iterations required to reach the optimal solution, while Figure 9(b) presents the lowest $\Delta\sigma$ values obtained for each $\lambda$ value (0, 0.5, and 1) across different number of initial simulations.

For the case of 100 simulations, the algorithm finds a low $\Delta\sigma$ value across all $\lambda$ values. With $\lambda = 0$, the algorithm achieves a $\Delta\sigma$ of 17.95 MPa after 55 iterations. Introducing $\lambda$ improves the performance further, with $\lambda = 0.5$ resulting in a $\Delta\sigma$ of 15.18 MPa and $\lambda = 1$ leading to the lowest value of 12.56 MPa after 66 iterations. These results suggest that adding an additional term helps improve the agreement between the simulated and experimental stress-strain curves owing to additional constraints.

When the initial number of simulations is reduced to 50, the BO algorithm still performs effectively, reaching $\Delta\sigma$ values less than to those obtained with 100 simulations. For $\lambda = 0$, the algorithm finds optimal parameters than yields a $\Delta\sigma$ of 11.99 MPa after 64 iterations, indicating a better performance than with 100 simulations. For $\lambda = 0.5$, the algorithm finds the optimum parameters more rapidly, reaching a low $\Delta\sigma$ of 12.09 MPa after 37 iterations, while $\lambda = 1$ results in a $\Delta\sigma$ of 18.86 MPa after 38 iterations.

With 25 initial simulations, the algorithm's performance decreases as expected, with higher Δσ values and more variability in the results. For $\lambda = 0$, the BO algorithm finds parameters that lead to Δσ of 29.78 MPa after 73 iterations, reflecting a larger error compared to larger initial number of simulations. With $\lambda = 0.5$, the algorithm reaches to minimum more quickly, achieving a Δσ of 23.55 MPa after 24 iterations, while for $\lambda = 1$, it takes 36 iterations to reach a Δσ of 36.89 MPa. These results indicate that with fewer initial simulations, the BO algorithm's ability to optimize the objective function becomes more limited, particularly when using $\lambda$ value, suggesting that with higher $\lambda$ values the contribution from the second term in the objective function is more significant leading to overall mismatch between stress-strain curve.

Finally, in the extreme case of zero initial simulation, the BO algorithm is entirely reliant on its exploration-exploitation balance to identify optimal parameter sets. As a result, the algorithm requires more iterations to find the minimum, with the lowest Δσ values being higher across all $\lambda$ values. For $\lambda = 0$, the algorithm achieves a Δσ of 23.5 MPa after 73 iterations, whereas $\lambda = 0.5$ leads to a higher Δσ of 49.75 MPa after 70 iterations and with $\lambda = 1$, the Δσ increases further to 61 MPa. This indicates that more iterations are needed to achieve better accuracy of the simulated stress-strain curve.

In summary, as the number of initial simulations decreases, the BO algorithm's ability to minimize Δσ is reduced. However, even with limited simulation data, the algorithm can still find reasonable solutions, particularly when $\lambda = 0.5$ is used with 50 initial simulations. As expected, starting with more simulations improves the algorithm's ability to explore and achieve a low Δσ, while introducing $\lambda$ helps control hardening behavior between simulated cycles, improving the accuracy of the simulated stress-strain response.

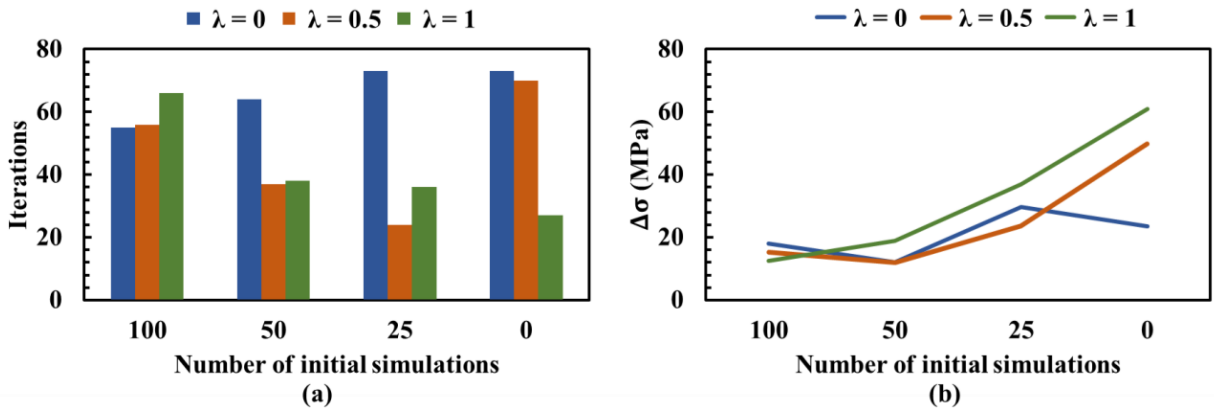

**Figure 9.** Performance of BO algorithm: (a) number of iterations to achieve optimal solution for different number of initial simulations with (b) corresponding Δσ values for each $\lambda$.

### 3.5 Optimization for multiple strains

After evaluating the algorithm for a single strain and determining that the combination of initial number of simulations of 50 and $\lambda = 0.5$ allows for efficient convergence to a low Δσ, the

BO algorithm is extended to optimize CP parameter two different strain amplitudes: ±0.5% and ±0.75% simultaneously. Since the optimization problem remains a single-objective function, Δσ is calculated as the average of the Δσ values obtained from the simulations at both ±0.5% and ±0.75% strain amplitudes. This approach ensures that the material response under different loading conditions is considered, leading to a more robust and generalized solution.

The algorithm performs effectively, achieving an average Δσ value of 15.06 MPa in 61 iterations, which indicates good agreement between the simulated and experimental stress-strain responses for both strain amplitudes. Figure 10 shows the stress-strain curves for both the ±0.5% and ±0.75% strain amplitudes. The simulations align well with the experimental data, indicating that the optimized parameters successfully capture the material behavior under both strain amplitudes.

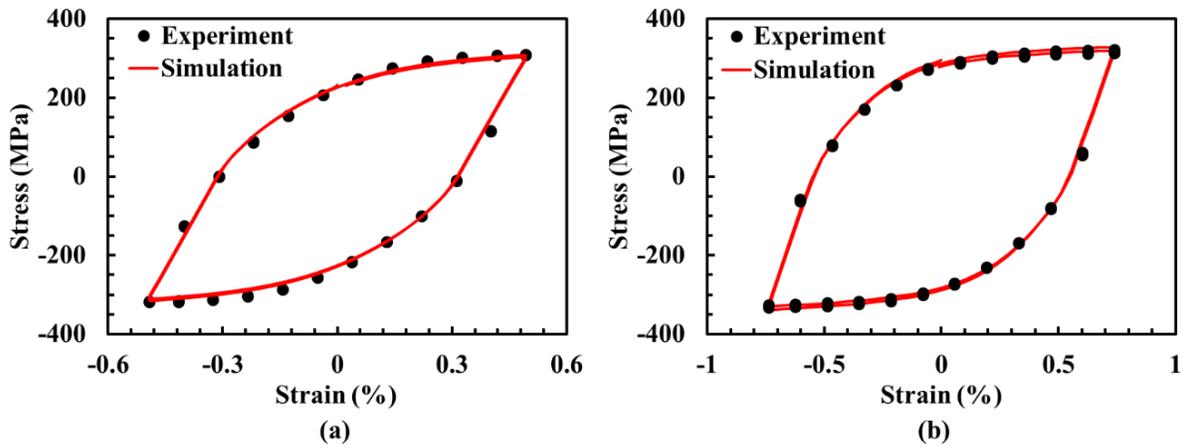

**Figure 10.** Stress-strain response corresponding to the parameters optimized by BO algorithm simultaneously for (a) 0.5% and (b) 0.75% strain amplitudes

By optimizing the parameters for different strain amplitudes, the BO algorithm demonstrates its capability to find optimum CP parameters across different loading conditions. The convergence to a low Δσ for both strain amplitudes indicates that the algorithm effectively balances exploration and exploitation. This results in the selection of parameter set that reduces the discrepancies between simulated and experimental stress-strain responses, ensuring the CP model performs well under varying loading conditions. The optimized parameters found using BO are tabulated in Table 3.

**Table 3.** Optimized parameters for the strain-gradient CP model for Hastelloy X at 500 °F.

| Parameter | Description | Optimized value |
| --- | --- | --- |
| $n$ | Strain rate sensitivity | 21.9 |
| $\tau_c^0$ | Initial CRSS (MPa) | 22.2 |
| $C$ | Geometric factor | 0.1 |
| $\rho_{SSD}^a$ | SSD density ($\mu m^{-2}$) | 83.7 |
| $h_0$ | Hardening rate (MPa) | 87.1 |
| $\tau_s$ | Saturation slip strength (MPa) | 437.4 |
| $m$ | Hardening exponent | 9 |
| $h$ | Kinematic hardening modulus (MPa) | 32694.6 |
| $h_d$ | Recovery modulus | 711 |

Figure 11 illustrates the evolution of the parameters during the optimization process across 75 iterations using the BO algorithm. Each subplot represents the trajectory of a specific material parameter, with the red dot indicating the final optimized value recommended by the algorithm. Initially, all parameters fluctuate, representing the exploration phase of the algorithm, before gradually stabilizing during the exploitation phase. In Figure 11(a), the strain rate sensitivity ($n$), exhibits a gradual stabilization, settling around 20 after about 20 iterations, indicating reliable convergence. Figure 11(b) shows that the initial CRSS ($\tau_c^0$) starts with fluctuations but eventually stabilizing around 20 MPa after 50 iterations. The geometric factor ($C$) in Figure 11(c), exhibits more variability but converges to a value of 0.1 by iteration 50, with the final optimized value aligning closely with the later iterations. Figure 11(d) depicts the evolution of the statistically stored dislocation density ($\rho_{SSD}^a$), which fluctuates between 40 and 120 before stabilizing around 80. Notably, $C$ and $\rho_{SSD}^a$ show coupled behavior - when one increases, the other decreases. This is expected, as the second term in Equation 5 is proportional to their multiplication which is shown in Figure 11(e). Initially, this combined parameter exhibits high variability but eventually converges around a value of 1, suggesting that the algorithm correctly understands their interaction.

The hardening rate ($h_0$) in Figure 11(f) shows significant variability throughout the iterations, reflecting the difficulty in identifying a stable value. However, most of the values lie in the range from 10 to 100 MPa, with occasional jump to a higher value. In Figure 11(g), the saturation slip strength ($\tau_s$) also starts with high variability but steadily decreases, converging around 450 MPa, indicating smooth optimization. Figure 11(h) shows the hardening exponent ($m$), which behaves consistently throughout the optimization process, converging around 9. Its smooth trajectory suggests that fewer iterations were needed for this parameter. The kinematic hardening modulus ($h$) in Figure 11(i) exhibits large fluctuations between 1,000 and 50,000 early on. However, by around 40 iterations, it stabilizes near 35,000. Finally, Figure 11(j) for recovery modulus ($h_d$) demonstrates high variability early in the optimization, but converges to a value near 700, consistent with the BO-recommended value.

Overall, the BO algorithm successfully narrows down the search for optimal parameter values. Parameters such as $n$, $\tau_s$, and $m$ exhibit smoother convergence trajectories, indicating they are more easily optimized. In contrast, parameters like $h_0$ and $h$ required more exploration. Additionally, the algorithm captures the interaction between $C$ and $\rho_{SSD}^a$. The final optimized values (marked by the red dots) align closely with the trends in the last few iterations, confirming the robustness of the BO process in identifying optimal parameters.

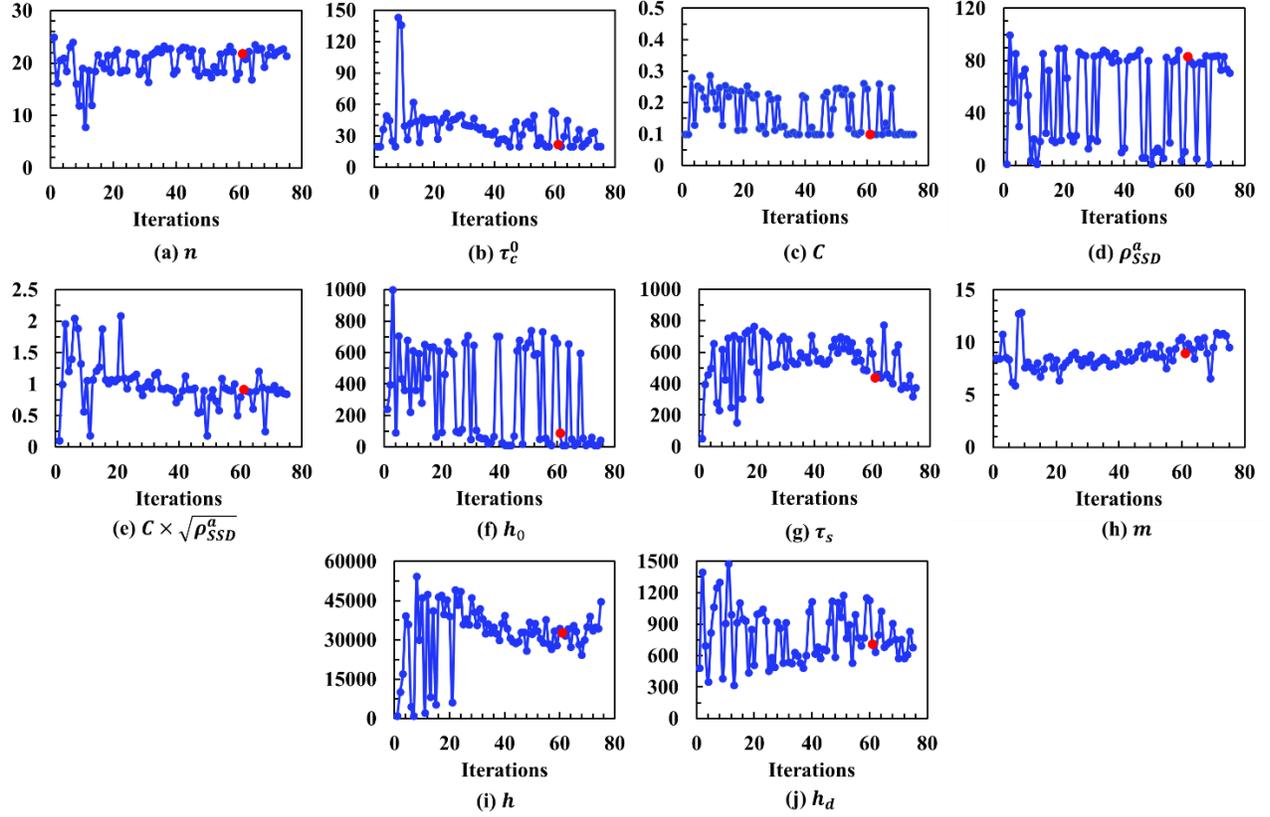

**Figure 11.** Trajectory of individual parameters (a) $n$, (b) $\tau_c^0$, (c) $C$, (d) $\rho_{SSD}^a$, (e) $C \times \sqrt{\rho_{SSD}^a}$, (f) $h_0$, (g) $\tau_s$, (h) $m$, (i) $h$ and (j) $h_d$ as suggested by BO algorithm over 75 iterations.

### 3.6 Sensitivity analysis

A sensitivity analysis is conducted to understand the influence of CP parameters on the stress at various points on the stress-strain curve. Figure 12 illustrates the impact of these parameters on different points of the stress-strain curve, highlighting how the contribution of each parameter varies across different stages of deformation. This variability reflects the underlying physical mechanisms governing material behavior. To explore these relationships, GP regression models are developed to quantify the effect of individual CP parameters on four key stress points along the stress-strain curve at: (1) the highest strain during tensile loading, (2) an intermediate strain during unloading, (3) the highest strain during compressive loading, and (4) an intermediate strain during tensile loading. The GP models achieve high accuracy, with an $R^2$ score of approximately 0.98 for testing data across all these locations. Such high $R^2$ values indicate that the GP models effectively capture the intricate relationships between the CP parameters and the stress-strain response, confirming the validity of the sensitivity analysis.

SHAP summary violin plots as shown in Figure 12 combines feature importance with the distribution of SHAP values for each parameter in the model. Each parameter is displayed on the y-axis, sorted by its overall importance, while the x-axis represents the SHAP value, indicating the parameter's positive or negative impact on the output (i.e., different points on stress-strain curve).

The plot's color gradient shows the feature (parameter) value (e.g., high or low), and the violin shape depicts the distribution of SHAP values for each feature across the dataset. The results reveal that the contribution of CP parameters varies at different stages of loading and unloading, offering insight into the material's behavior. At the highest strain during tensile loading, the parameters $\rho_{SSD}^a$, $\tau_c^0$ and $C$, show the most significant influence on the stress response (Figure 12(a)). According to Equation 5, these parameters significantly affect the material's yield point [2], which directly impacts the stress level at maximum strain. Additionally, the hardening modulus $h$,

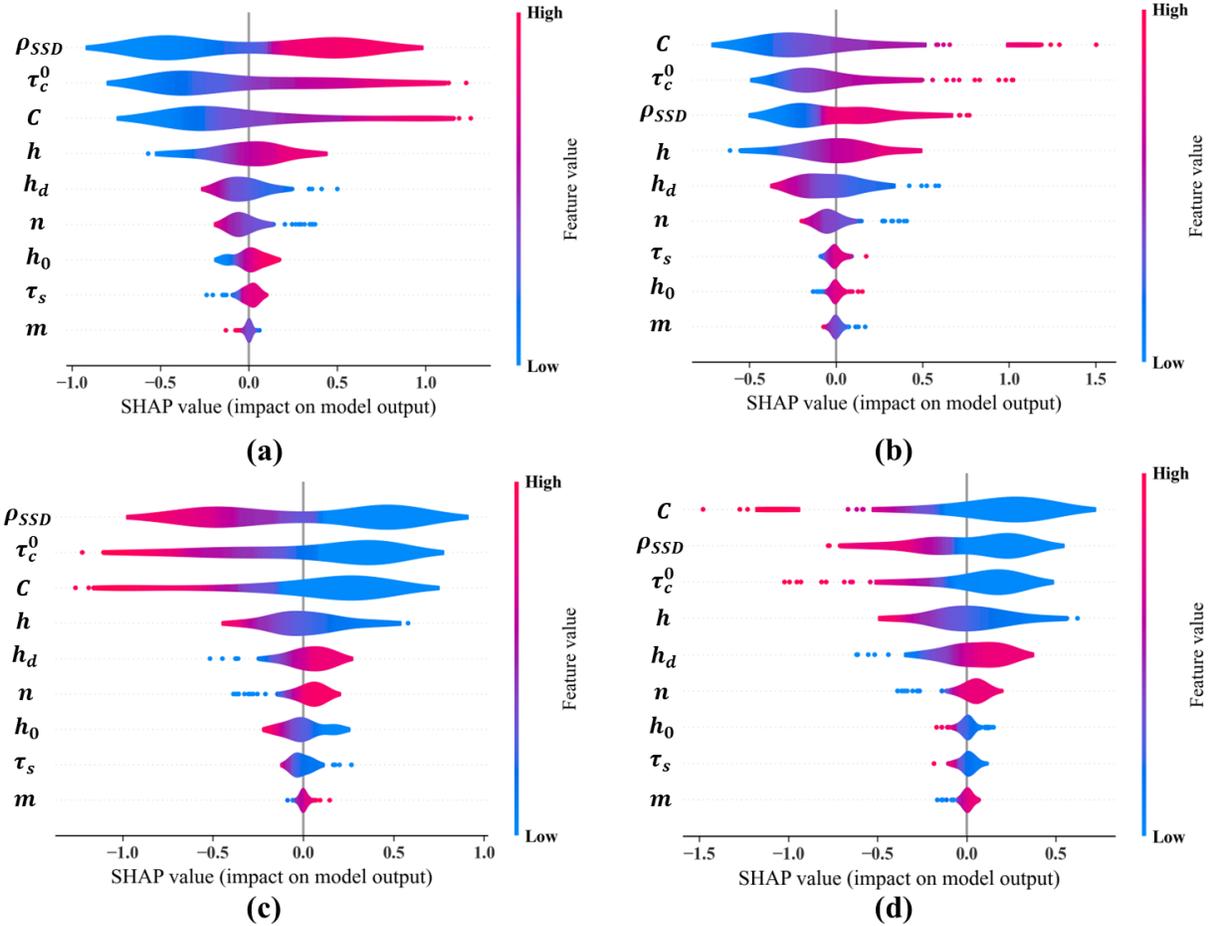

**Figure 12.** SHAP summary plot for influence of CP parameters on stress at (a) the highest strain during tensile loading, (b) the intermediate strain during unloading, (c) the highest strain during compressive loading and (d) the intermediate strain during tensile loading

representing kinematic hardening, plays a key role, contributing to the stress at maximum strain. The remaining parameters have secondary effects, indicating that while they influence the stress-strain behavior, their impact is less pronounced in comparison to the primary factors. Moreover, for the top four parameters in this plot it is observed that the stress increases (positive impact on model output) as the values of these parameters increase (low to high from left to right). For the intermediate strain during unloading, a similar pattern is observed in Figure 12(b), with $C$, $\tau_c^0$ and

$\rho_{SSD}^a$ remaining dominant. These parameters control the material's yield point, leading to an upward or downward shift in the entire stress-strain curve after the onset of plastic deformation [57]. However, in this case, the recovery modulus $h_d$, associated with backstress, shows a greater spread, indicating that stress is more sensitive to changes in this parameter. This suggests that the evolution of backstress plays a more prominent role during unloading than during tensile loading [22]. At the highest strain during compressive loading (Figure 12(c)), the contribution is very similar to the highest strain level during tensile loading, with $\rho_{SSD}^a$, $\tau_c^0$ and $C$ dominating the stress response. Finally, in Figure 12(d), corresponding to intermediate strain during tensile loading, the observed behavior closely matches the intermediate strain during unloading. However, the hardening parameter $h_0$ begins to show a stronger effect as compared to unloading stage, likely indicating the onset of strain hardening as the material prepares to undergo further deformation.

In addition, a separate GP regression model is trained to understand the impact of CP parameters on the difference in stress at the end strain levels, i.e., $\Delta\sigma_{max32}$ and $|\Delta\sigma_{min32}|$. $R^2$ score for both the GP models is approximately 0.96. In Figure 13(a), it is evident that hardening parameters, particularly $h_0$ (hardening rate) contribute substantially to the observed stress difference $\Delta\sigma_{max32}$. This is consistent with the established behavior of materials under deformation, wherein an elevated hardening rate typically enhances resistance to deformation, thereby resulting in more pronounced difference in stress between consecutive cycles [57]. Similar observations are made from Figure 13(b) as well, with hardening parameters dominating the response. However, increased spread is observed from backstress parameters ($h$ and $h_d$) during compressive loading. This indicates these parameters have greater influence on the output during compressive loading as compared to during tensile loading. This observation aligns with the physics of deformation, where backstress evolution due to dislocation interactions is observed during compressive loading that result in more pronounced kinematic hardening effects [22].

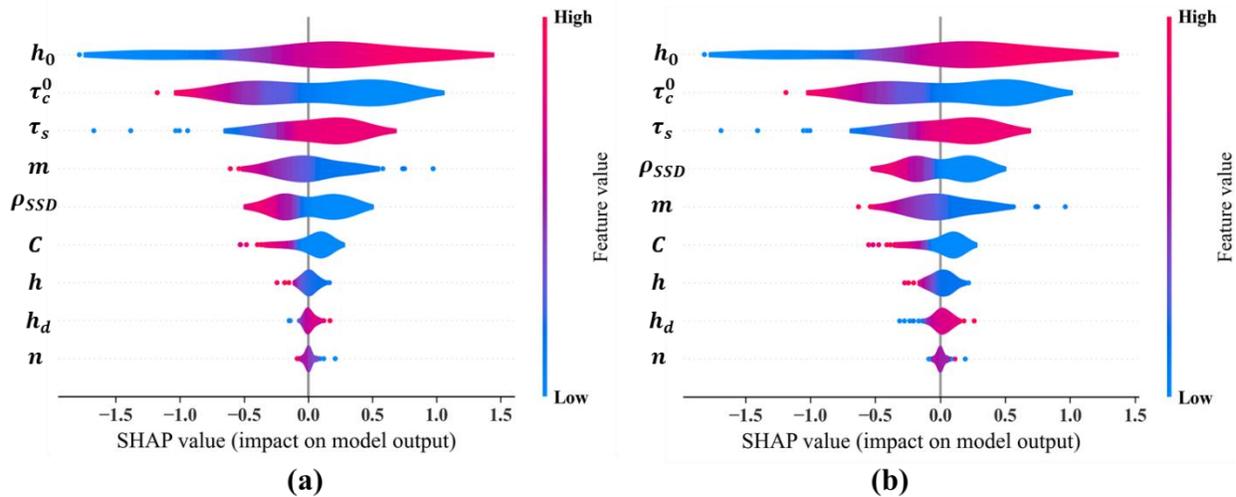

**Figure 13.** SHAP summary plot for influence of CP parameters on the difference in stress at the end levels during (a) tensile loading ($\Delta\sigma_{max32}$) and (b) compressive loading ($|\Delta\sigma_{min32}|$)

Overall, this sensitivity analysis identifies the most influential CP parameters at various points along the stress-strain curve, enabling improved control and prediction of material behavior during different stages of deformation. The results provide a clear pathway for refining experimental efforts by prioritizing the measurement of parameters with the greatest impact on material response. Such targeted experimentation enhances the fidelity of the CP model, ultimately leading to more accurate and reliable predictions of material behavior under varying loading conditions.

## 4. Discussion

### 4.1 Comparison of mechanical response of RVE with and without twins

Twins are observed in L-PBF Hastelloy X as shown in Figure 1(b), however, previous work [35] for Hastelloy X did not include twins in the synthetic microstructure. Therefore, first, a synthetic microstructure is generated without twins, as shown in Figure 14(a). Then, twins are inserted in the same microstructure, as shown in Figure 14(b), using the method described in Section 2.2.1. The colors of the grains in the microstructure correspond to the number assigned to them. A comparison between the macro- and micromechanical response of the synthetic microstructure with and without twins is conducted to understand the impact of inserting twins during simulations. These simulations use a finer mesh (3 µm), resulting in approximately 1000 elements per grain. The analysis, detailed in the following sections, is based on simulations run for 5 loading cycles using optimized parameters as listed in Table 3. Figure 14(c) presents the homogenized macroscopic stress-strain curves of the synthetic microstructures with (red line) and without twins (black dots). Interestingly, the overall stress-strain response for both cases is nearly the same. This suggests that the introduction of twins in the synthetic microstructure does not significantly influence the homogenized macroscopic mechanical response under the applied loading conditions. However, the likely location of failure identified using the highest value of $W$ shifts from a normal GB in the RVE without twins to a TB after the insertion of twins. This suggests that while twins may not contribute significantly to overall mechanical response, their presence alters the pathways of local deformation, affecting dislocation motion and concentrating strain energy at TBs [58].

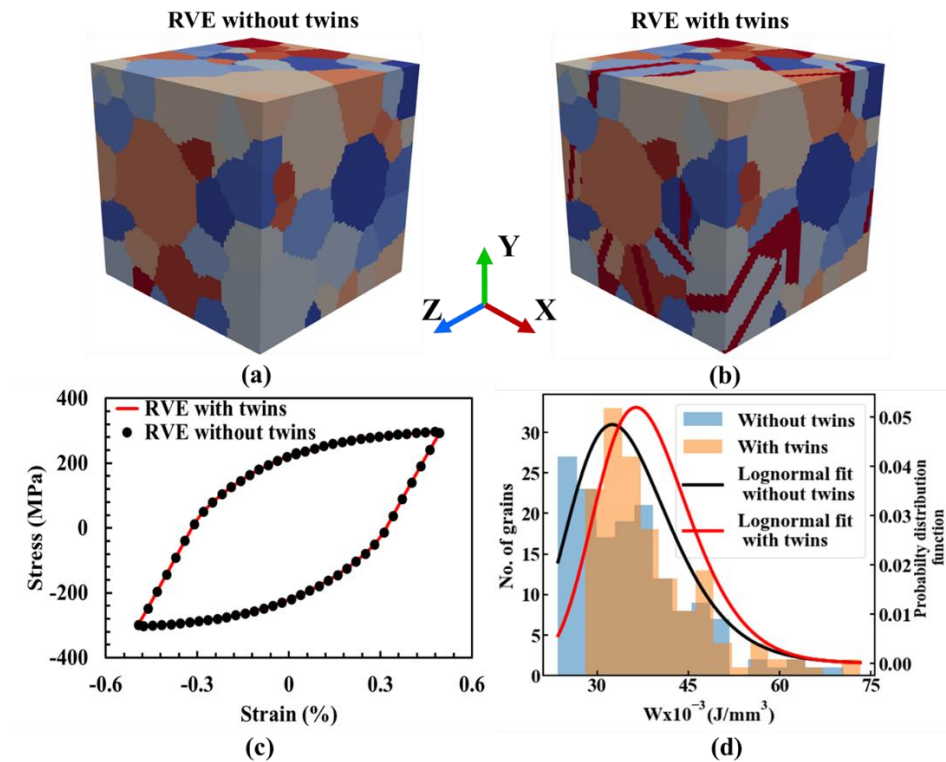

**Figure 14.** RVE (a) without twins and (b) with twins and corresponding (c) stress-strain response when subjected to 0.5% strain amplitude. (d) Distribution of strain energy density of top 150 grains for RVE with and without twins.

To further explore this change in the micromechanical response, the distribution of maximum accumulated $W$ is plotted for the top 150 grains from both microstructures. Figure 14(d) shows histograms of accumulated strain energy density in each microstructure. Blue bars and orange bars represent data for microstructure without and with twins, respectively. The lognormal fits highlight a rightward shift for the microstructure with twins (red line) compared to the fit for microstructure without twins (black line), indicating that twins enhance the accumulation of strain energy in many grains. This shift in the distribution suggests that the inclusion of twins increases the accumulation of strain energy density in the microstructure. These findings emphasize the critical role of twins in altering the micromechanical response and enhancing strain energy accumulation, thereby impacting the material's fatigue behavior and failure likelihood.

### 4.2 Effect of twins on the micromechanical fields

Figure 15(a) illustrates a cross-section of a representative RVE containing twins, along with the corresponding von Mises stress contour under fatigue loading conditions. In Figure 15(b), the von Mises stress contour map corresponding to the same RVE cross-section highlights how the stress is distributed across the grains when subjected to load in the Z-direction. The color map reveals significant stress variations, with stress accumulation predominantly occurring around GBs and TBs. These high stress regions are critical, as they may serve as initiation sites for fatigue damage or crack nucleation. Notably, the von Mises stress appears to be high near these boundaries, confirming the role of boundaries as stress concentrators [31]. This heterogeneous stress distribution underscores the influence of microstructural features, such as grain orientation, size, and the presence of twins, on the overall stress state.

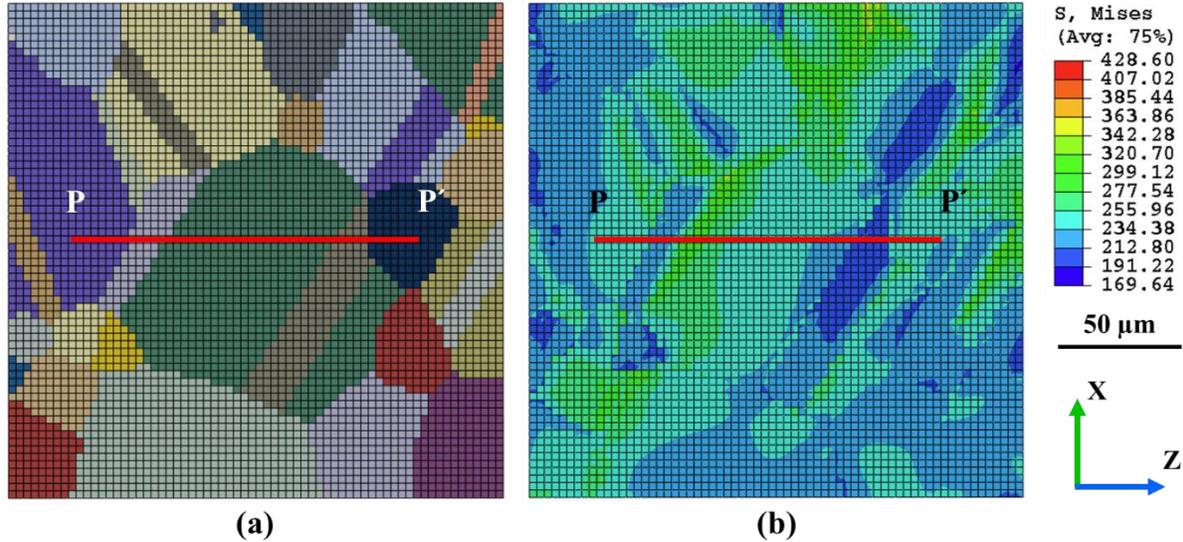

**Figure 15.** Cross section of (a) a representative RVE with twins and (b) the corresponding von Mises stress contour.

Slip is the primary mechanism for plastic deformation in polycrystalline materials during fatigue [31], [59]. As slip accumulates during cyclic loading, strain localization and accumulation of strain energy are observed, which lead to crack initiation. When the residual shear stress (RSS) exceeds the critical resolved shear stress (CRSS), i.e. when the ratio of RSS to CRSS is greater than 1, according to Equation 4, slip initiates within the material. Slip continues along the path of low energy which is preferred by dislocations. As the dislocations begin to glide on the activated slip system, they encounter numerous obstacles, such as TBs or GBs. This leads to dislocation pile-ups at these boundaries, resulting in stress concentration. In low stacking fault energy materials, such as Hastelloy X, TBs accommodate a significant portion of the plastic deformation. This accumulation of plastic strain and stress concentration at the TBs ultimately leads to crack nucleation, indicated by high stored strain energy density [60].

Figure 16 provides detailed insights into various micromechanical attributes along the path PP', shown in Figure 15, focusing on their variations as affected by the presence of GBs and TBs in a microstructure under cyclic loading conditions. This analysis is based on the results obtained after the 5th loading cycle, which highlights the changes in von Mises stress, plastic strain, and other critical physical attributes as a function of normalized distance across the microstructure. In Figure 16(a), the maximum ratio of RSS to CRSS is plotted. This ratio peaks at both GBs and TBs, indicating regions where plasticity is most pronounced due to high local stress, which is consistent with literature highlighting high RSS values at TBs due to elastic incompatibility stresses arising from crystallographic orientation differences [61]. The net effect, including external loading stress and incompatibility stress, can be evaluated using the von Mises stress. Figure 16(b) shows the von Mises stress distribution along the same path, with notable peaks at TBs and GBs, suggesting elevated stress concentrations at these locations. Corresponding to high RSS values, the slip accumulation and high strain gradients leading to higher dislocation densities along the TBs can be observed in Figure 16(c) and (d), respectively. Higher slip accumulation means a pile-up of dislocations in the vicinity of the TBs, leading to subsequent plastic strain localization as depicted in Figure 16(e). Finally, Figure 16(f) presents the strain energy density along the same path, with elevated values at TBs and GBs as a result of slip localization and high strain gradients. This higher energy concentration suggests that TBs are likely sites for crack nucleation and eventual failure, driven by the accumulation of strain and dislocation activity.

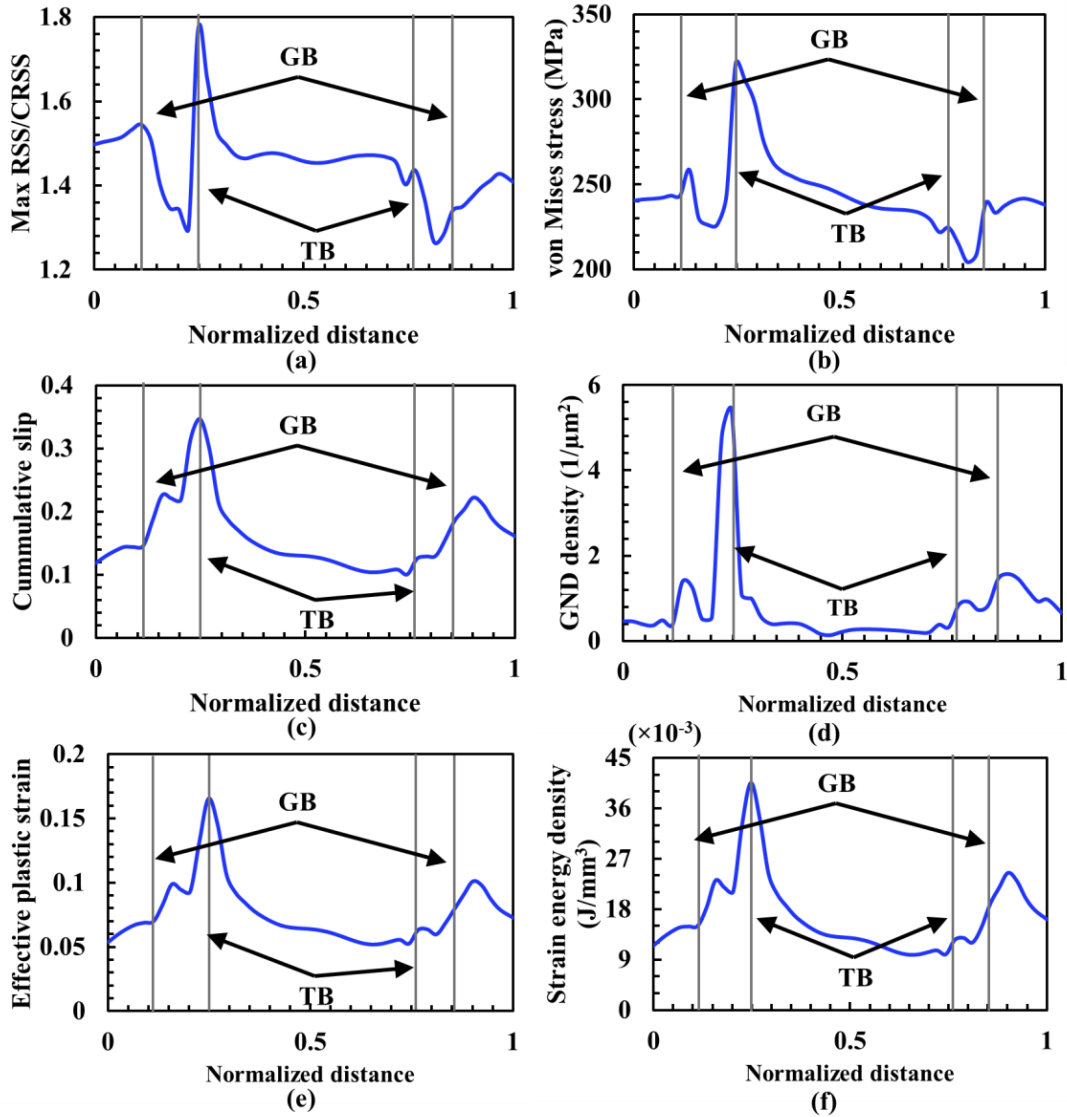

**Figure 16.** Line plot across path PP′ for (a) maximum ratio of RSS to CRSS, (b) von mises stress, (c) cumulative slip, (d) GND density, (e) strain localization, and (f) strain energy density.

In summary, Figure 16 highlights the critical role of TBs in controlling the micromechanical behavior of polycrystalline materials under cyclic loading. The presence of TBs leads to localized plastic deformation, higher dislocation density, and elevated strain energy density, all of which contribute to the material's potential for fatigue damage and failure. These results align with the understanding that TBs can act as preferential sites for crack initiation due to the strain localization and accumulation of strain energy density [62], [63].

**4.3 Identification of factors influencing damage via plastic strain energy density**

As discussed above, strain energy density can be used to indicate potential fatigue failure initiation sites. To determine the potential location of crack initiation during high-temperature fatigue loading conditions, the highest value of $W$ is identified for each grain in every RVE. Upon inspection of the $W$ field across all 10 RVEs, it is observed that the highest value of $W$ is typically found at the TBs, as listed in Table 4.

Accumulated damage typically results from dislocation generation and accumulation. As dislocations multiply and interact, they contribute to strain hardening and microstructural changes, which can eventually lead to the initiation of microcracks [31]. From the results, it is observed that extreme accumulated damage, as indicated by dissipative energy per unit volume ($W$), is related to high gradients in plasticity, i.e., GND density. A quantitative analysis of $W$ and GND density is performed using highest value of $W$ from each grain in the RVE for all 10 RVEs. Cumulative distribution functions (CDF) of GND density are generated to analyze the relationship between accumulated damage and high gradients in plasticity, as shown for representative RVEs in Figure 17. The GND density displayed is a summation of the GND density from all 12 slip systems. The red dot in Figure 17(a) and (b) represents the value of GND density for the grain corresponding to the location of the maximum accumulated damage, as predicted by $W$. The CDF value of GND density, except for RVE#8 is generally greater than 0.9, indicating the extreme value of accumulated damage ($W$) associated with high gradients in plasticity.

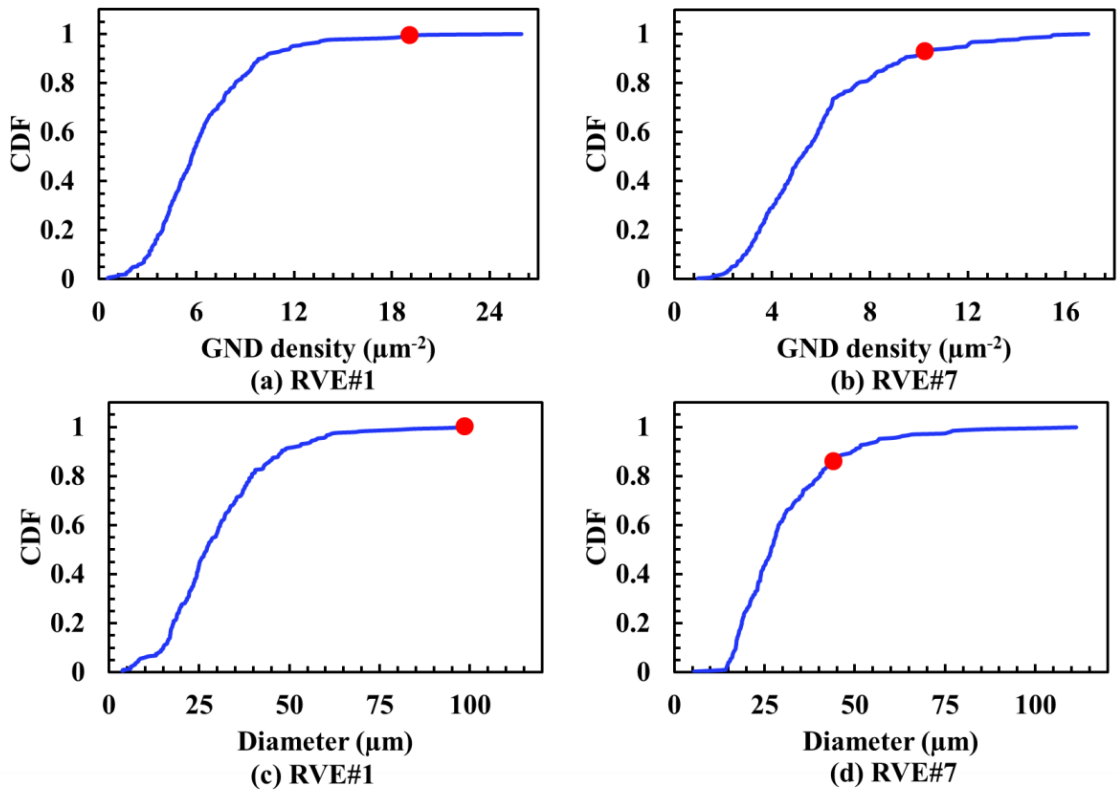

**Figure 17.** Representative cumulative distribution function plots of (a), (b) GND density and (c), (d) equivalent diameter of the grains for RVE#1 and 7, respectively. The red dot represents the grain corresponding to the highest accumulated $W$.

Similar CDF plots, Figure 17(c) and (d), are generated for the equivalent sphere diameter using data from each RVE. Additionally, values of the Schmid factor and average misorientation of the grains are recorded at the likely failure locations. The average misorientation of a grain is the number-weighted average of the misorientation with neighboring grains, providing insights into the grain's interaction with its neighbors. It is observed that the location of failure is associated with the grain size. The CDF value at the location of the highest accumulated $W$ is greater than 0.8 for all the cases except RVE#8, as listed in Table 4. This suggests that failure is highly likely to occur in the larger grains within the microstructure. This is consistent with the literature [31], where it was concluded that larger grains were more prone to cracking. The Schmid factor for these grains with the highest $W$ is generally high, typically around 0.46, indicating that the grain is favorably oriented with the loading direction, as the maximum possible value for the Schmid factor is 0.5. Furthermore, the average misorientation for these critical grains is observed to range between 39° and 45°. These observations suggest that failure is not due to extreme conditions, rather it occurs when multiple factors align themselves such as larger grains that are favorably oriented (high Schmid factor) and are neighboring grains with high misorientation.

**Table 4.** List of failure locations with corresponding cumulative distribution function values for GND density and diameter, along with the average misorientation and Schmid factor of grains exhibiting the highest accumulated plastic strain energy density for each RVE.

| RVE# | Twin boundary | GND density CDF value | Diameter CDF value | Average misorientation | Schmid factor |
|---|---|---|---|---|---|
| 1 | No | 0.99 | 1 | 42.89 | 0.49 |
| 2 | Yes | 0.94 | 0.97 | 41.63 | 0.41 |
| 3 | No | 0.99 | 0.8 | 43.18 | 0.45 |
| 4 | Yes | 0.9 | 0.89 | 43.25 | 0.48 |
| 5 | Yes | 0.99 | 0.86 | 39.12 | 0.48 |
| 6 | Yes | 0.93 | 0.86 | 43 | 0.48 |
| 7 | No | 0.92 | 0.97 | 44.04 | 0.49 |
| 8 | Yes | 0.78 | 0.71 | 45.25 | 0.39 |
| 9 | Yes | 0.96 | 0.96 | 41.25 | 0.43 |
| 10 | Yes | 0.94 | 0.9 | 42.6 | 0.48 |

To summarize, a relationship has been observed between the plastic strain energy density (fatigue indicator parameter) and microstructural features such as grain diameter, average misorientation, and Schmid factor. The analysis reveals that the highest damage as indicated by high strain energy accumulation typically occurs in grains with diameters at the upper end of the distribution and near the TBs. These grains are favorably oriented, as indicated by high Schmid factors, and exhibit an average misorientation of approximately 42°±1.67°.

## 5. Conclusion

This study presents a systematic approach for calibrating strain gradient crystal plasticity (CP) model parameters using Bayesian optimization (BO), while evaluating the influence of CP parameters on the material's mechanical response through sensitivity analysis. The results demonstrate that the BO framework effectively identifies optimal CP parameters with minimal computational effort, significantly reducing the error between experimental and simulated stress-strain curves. Additionally, the impact of introducing twins in microstructure sensitive modeling is analyzed. Furthermore, using quantitative approach key factors that drive fatigue damage in Hastelloy X are identified. The key findings are:

- The Gaussian process surrogate model demonstrated strong predictive capability with R² scores around 0.92, enabling efficient optimization and faster convergence. Using the BO algorithm, the CP model with as few as 50 initial simulations, finding optimal parameters within 75 iterations across two different strain levels. An additional term in the objective function improved the fit between experimental and simulated curves by controlling the hardening effect between cycles, although excessive weighting of this term could negatively affect overall fit.
- Sensitivity analysis revealed that parameters $\rho_{SSD}^\alpha$, $\tau_c^0$ and $C$ have a dominant impact on the stress-strain response at various load points, with increased influence from backstress parameters during compressive loading. Isotropic hardening parameters are found to dominate when evaluating the difference between two consecutive loading cycles.
- Introducing twins into the synthetic microstructure affects the micro-scale response while the macro-scale stress-strain response remains unaltered. Also, the likely location of failure shifted from a normal grain boundary to a twin boundary while increasing accumulated strain energy density in many grains in the microstructure.
- The analysis using CDF revealed a relationship between fatigue damage and microstructural features, identifying larger diameter grains with higher Schmid factors and average misorientation of approximately 42°±1.67° as probable failure sites.

Overall, the integration of GP with BO and sensitivity analysis provides a powerful framework for calibrating CP models and understanding the influence of individual parameters on the material behavior. Future research could expand the calibration process to include multiple objective functions, allowing for improved control over specific points on the stress-strain curve rather than averaging across the entire curve, as done in this study. Furthermore, insights gained from the sensitivity analysis could assist in assigning appropriate weights to parameters or reducing the parameter set by identifying those that are least influential. Recently, Graph Neural Networks have garnered attention in various predictive applications [64], [65], [66] and could be utilized as surrogate models to rapidly predict mechanical response of complex material behaviors, such as fatigue damage evolution. Moreover, the findings regarding the impact of microstructural features, such as TBs and grain characteristics on fatigue behavior, highlight the potential for tailoring materials to enhance performance and reliability under cyclic loading.

**CRediT authorship contribution statement**
**Ajay Kushwaha:** Writing – original draft, Visualization, Validation, Software, Methodology, Investigation, Formal analysis, Data curation. **Eralp Demir:** Writing – review & editing, Methodology, Software. **Amrita Basak:** Writing – review & editing, Supervision, Resources, Project administration, Funding acquisition, Conceptualization.

**Declaration of competing interest**
The authors declare that they have no known competing financial interests or personal relationships that could have appeared to influence the work reported in this paper.

**Declaration of generative AI in scientific writing**
During the preparation of this work, the first author, Ajay Kushwaha, used OpenAI ChatGPT in order to improve grammar and readability. After using this tool/service, the first author reviewed and edited the content as needed and takes full responsibility for the content of the publication.

**Acknowledgements**

The authors would like to thank Solar Turbines Incorporated (San Diego, USA) for providing the Hastelloy X specimens used for EBSD characterization. Authors would also like to thank Dr. Darren Pagan, Assistant Professor, Materials Science and Engineering and Mechanical Engineering at Penn State for valuable discussions regarding crystal plasticity simulations.

**Funding information**

This work is supported in part by Solar Turbines Incorporated, a Caterpillar Company through grant number PAT-18-00112, in part by the US Department of Energy's Nuclear Engineering University Program (NEUP) through grant number DE-NE0009253, and in part by the Department of Mechanical Engineering at Penn State. Any opinions, findings, and conclusions in this paper are those of the authors and do not necessarily reflect the views of the supporting institution.

**Data availability**
Data will be made available on request.